# Upscaling the hyperpolarization sample volume of an automated hydrogenative parahydrogen-induced polarizer

Yenal Gökpek[a], Jan-Bernd Hövener*[a], Andrey N. Pravdivtsev*[a]

[a] Section Biomedical Imaging, Molecular Imaging North Competence Center (MOIN CC), Department of Radiology and Neuroradiology, University Hospital Schleswig-Holstein, Kiel University, Am Botanischen Garten 14, 24114, Kiel, Germany

**ABSTRACT:** Nuclear magnetic resonance (NMR) and magnetic resonance imaging (MRI) suffer from inherently low sensitivity due to the weak thermal polarization of nuclear spins. Parahydrogen-induced polarization (PHIP) offers a powerful route to enhance NMR signals by several orders of magnitude, enabling real-time metabolic imaging. However, PHIP implementations are often constrained by small sample volumes, limited automation, and complex high-pressure requirements. In this work, we present an upgraded, automated PHIP system capable of hyperpolarizing sample volumes up to 2.2 mL, suitable for preclinical MRI applications. We developed several high-pressure reactors and multi-port NMR tube caps compatible with standard commercial 5, 10, and 16 mm glass tubes. Reactor designs were simulated and fabricated from chemically resistant polymers, ensuring mechanical safety at more than 30 bar. Using FLASH MRI, nutation, and CPMG sequences, we characterized magnetic field homogeneity and stability, establishing optimal sample dimensions (12.5/16 mm ID/OD glass tube, 20 mm height) with $B_0$ inhomogeneity below 2.5 ppm and $B_1$ inhomogeneity around 1%. A high level of injection reproducibility was confirmed (volume precision ~0.6%). Optimization of experimental parameters, including hydrogenation pressure, pH$_2$ flow rate, and sample temperature, enabled rapid and efficient polarization transfer. At optimized conditions (20 bar pH$_2$, 2 L/min flow, 55°C, 4 s bubbling time), up to 31.3% $^1$H polarization of two protons was achieved for deuterated ethyl acetate in acetone with the theoretical maximum of 50%. This level of polarization was achieved with a duty cycle of 80 s, and the standard deviation of the mean was below 6.8%. This system lays the groundwork for the broader adoption of PHIP in preclinical imaging and metabolic research, providing practical sample volumes and facilitating the rapid production of hyperpolarization. Future work includes automating the purification process and further maximizing the polarization yield.

## Introduction

Nuclear magnetic resonance (NMR) uses weak nuclear spin interactions with the magnetic field to gain insights into molecular structure and sample composition[1]. Combined with magnetic field gradients, it enables noninvasive magnetic resonance imaging (MRI)[2]. However, the sensitivity of NMR is low, partially because of the low nuclear spin polarization: at fields of 9.4 T and 300 K, the polarization of $^1$H is about $3.2 \times 10^{-5}$. Thus, even at 9.4 T, the MR signal can be enhanced up to 31.000 times when unity polarization is reached. Consequently, much effort is spent on researching methods to increase the polarization[3,4]. Dissolution dynamic nuclear polarization (DNP)[5,6], parahydrogen-induced nuclear polarization (PHIP)[4,7–9], and signal amplification by reversible exchange (SABRE)[10] are among the most popular methods for hyperpolarizing molecules in solution. Remarkably, hyperpolarized MRI has enabled real-time noninvasive observation of metabolic transformations in vitro[11–13] and in vivo[14,15].

Hydrogenative PHIP typically utilizes a precursor with an unsaturated double or triple C-C bond, parahydrogen (pH$_2$), and a catalyst to promote the hydrogenative reaction[16–18]. pH$_2$ is a lower-energy, singlet nuclear spin isomer of dihydrogen with a total nuclear spin of 0. It can be conveniently produced by cooling H$_2$ gas to 77 K, where pH$_2$ is enriched to ~50% [19,20] or 25 K and below, where the enrichment is close to 100%[21,22].

To polarize biomolecules like pyruvate and acetate, which don't have a native unsaturated precursor, an unsaturated side arm can be added to receive pH$_2$. After hydrogenation, the polarization can be transferred to the target, and the side-arm cleaved (**Figure 1**). This side-arm hydrogenation (SAH) method (PHIP-SAH) was first proposed by Reineri et al.[23,24]. Its theory and practices were recently reviewed by Salnikov et al.[25].

Most of the PHIP-SAH studies in the literature were conducted with 5 mm NMR tubes and small volumes (0.1-0.2 mL)[26–28], which is reasonable in terms of resources. However, there are examples when larger-volume reaction vessels are constructed[29,30], or 10 mm NMR tubes are used[31,32].

In general, PHIP experiments are rather complex and require well-defined and synchronized events of chemistry, fluidics, and NMR. Thus, since the beginning, a key element for well-defined and reproducible hyperpolarization experiments has been automation and quality control[33–36]. For example, Schmidt et al.[37] demonstrated a semi-automated approach in which a 0.7 mL sample was injected, infused with pH$_2$, and ejected without purification every 15 s.

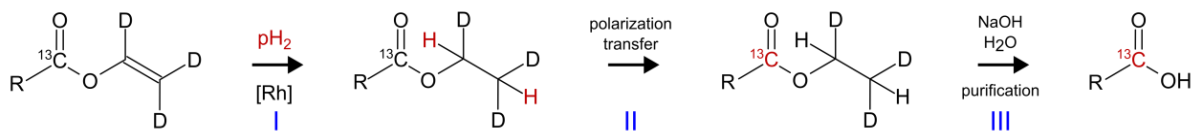

**Figure 1. Schematic of the PHIP-SAH experiment.** First, $^{13}$C-labeled deuterated vinyl ester with a residue group (noted by R; acetate for R = CD$_3$, pyruvate for R = COCD$_3$) is hydrogenated to ethyl ester (I). Then, the pH$_2$ spin order is converted into $^{13}$C spin magnetization (II). Finally, after the addition of NaOH to the aqueous solution to promote the cleavage of the side arm and purification, the $^{13}$C hyperpolarized ester can be extracted (III). In this work, we used vinyl acetate (VA) precursor, which upon hydrogenation yielded ethyl acetate (EA). Protonated (VA-d$_6$) and deuterated (VA-h$_6$) precursors were compared.

A commercial polarizer prototype for PHIP-SAH was also used by Nagel et al. for the preparation of nuclear spin hyperpolarization and consequent in vivo imaging[38].

Recently, we introduced a semi-automated, PHIP-SAH compatible polarizer with a duty cycle of about 1 minute, based on a permanent magnet portable MRI unit[39]. The system operated with 10 mm NMR tubes at pH$_2$ pressures up to 30 bar. Larger samples suitable for animal imaging, however, were not feasible. Thus, we set out to develop a setup suitable for larger sample sizes in preclinical imaging.

Here, we present a novel polarizer and its performance for polarizing samples up to 2 mL, suitable for small animal imaging[6,38,40].

First, we aim to maximize the sample size (more than 2 mL) by varying the sample tube diameter (**Figure 2**) and sample height (**Figure 3**), while keeping the homogeneity of $B_0$ (less than 5 ppm) and $B_1$ across the sample. To achieve this goal, we developed and tested several PHIP reaction chambers based on high-pressure multiport NMR tube caps, 5- and 10-mm NMR tubes, and 16 mm microwave tubes. High $B_0$ and $B_1$ homogeneity throughout the entire sample is needed for a robust spin order transfer (SOT)[31]. On the other hand, automation enabled us to perform precise serial experiments, for example, by measuring the hydrogenation kinetics of vinyl acetate at various temperatures. After several iterations, we achieved a maximum absolute $^1$H polarization of 31.3% for two protons, out of a possible 44.6% for the used pH$_2$ enrichment level. We believe that this is an essential step toward maximizing the hyperpolarization yield of automated hydrogenative PHIP.

**Results**

We describe the characterization of the conditions under which PHIP was performed, including the design of reaction vessels and the optimization of hydrogenation parameters and SOT. Through these optimizations, we achieved a $^1$H net polarization of 31.3% for ethyl acetate-d$_6$, close to the theoretical maximum of 50% for the selected SOT scheme. The duty cycle of the automated polarization protocol was 80 s.

**$B_0$ field homogeneity: sample diameter**

The $B_0$ homogeneity was evaluated in 5, 10 mm NMR tubes and a 16 mm OD microwave tube filled to different amounts (**Figure 2**, inner diameters (ID) 4.2 mm, 9.1 mm, and 12.5 mm). For 20 mm filling, non-localized $^1$H NMR of acetone yielded full widths at half maximum (FWHM) of 1.047, 1.574, and 2.492 ppm (19.855, 29.847, and 47.254 Hz), for 5, 10, and 16 mm tubes, respectively.

In situ $^1$H FLASH MRI[2] (**Figure 2a-c**) showed homogeneous images of the samples, indicating a sufficiently homogeneous $B_0$ magnetic field across the sample. Thus, we continued to investigate the 16 mm reactor, which provides the desired volumes.

**$B_0$ and $B_1$ field homogeneity: sample height**

We filled the 16 mm reactor to heights, $h$, of 10, 20, 30, and 40 mm, and measured $^1$H spectra, $^1$H FLASH, and nutation curves (**Figure 3**, note that the FLASH images represent a combination of $B_0$ and $B_1$, while nutation curves assess only $B_1$ homogeneity). The images were homogeneous for up to 20 mm filling and showed artefacts for $h$ > 20 mm.

Fitting of nutation curves with a damped sine wave yielded decay rates of 57.8 ± 3.4, 369.7 ± 23.4, 521.8 ± 61.4, and 531.9 ± 76.6 s$^{-1}$ for samples with 10, 20, 30, and 40 mm heights, respectively. Thus, the $B_1$ homogeneities of the $h$ = 30 and 40 mm samples were very similar. The nutation period was measured to be 90.30 ± 0.14 μs for a height of 20 mm. It could be seen that $B_1$ homogeneity decreases with increasing sample height; however, the signal decay for the $h$ = 20 mm sample after 5 nutation periods is only 6.7%.

To estimate the $B_1$ field homogeneity more precisely, we simulated nutation curves as a decaying sine function, assuming a Gaussian distribution of the $B_1$ field across the sample, and tried to determine the $B_1$ field homogeneity from this simulation in terms of standard deviation. We obtained the same nutation decay parameters as those experimentally obtained (SI, **Figure S1**). The resulting simulations yielded the following estimates for the relative standard deviations of the $B_1$ fields: approximately 0.26%, 1.03%, 2.30%, and 4.08% for $h$ = 10, 20, 30, and 40 mm, respectively. Hence, we concluded that the 20 mm bore magnet creates a sufficiently homogenous magnetic field in the center of the field of view (FOV) with a 20 mm height and 12.5 mm width. These measurements indicated that the system is suitable for reactors with a sample size of (2.215 ± 0.0140) mL, providing $B_1$ inhomogeneity of about 1% and $B_0$ inhomogeneity in $^1$H frequencies better than 47 Hz (2.492 ppm).

**$B_0$ field stability: measurement of $T_2$ relaxation time**

The $B_1$ and $B_0$ stability and homogeneity inside the 5, 10, and 16 mm reactors were measured with a Carr-Purcell-Meiboom-Gill (CPMG)[41,42] sequence (**Figure 4**). The CPMG experiments were repeated using different echo time intervals, $2\tau$, from 2 to 800 ms, where $2\tau$ is the time between two consecutive refocusing pulses. To ameliorate the effects of $B_1$ inhomogeneity, we used composite pulses, $90°_Y 180°_X 90°_Y$, instead of a single inversion pulse[31].

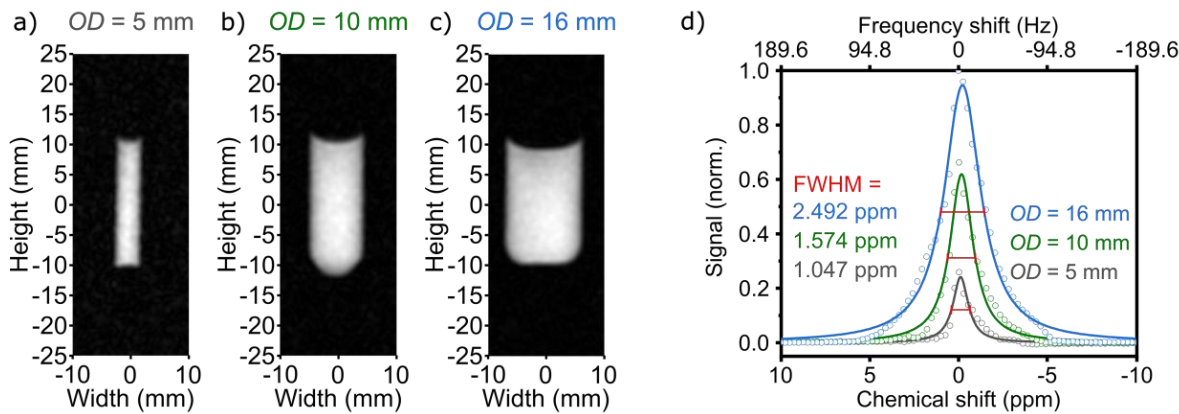

**Figure 2. Effect of sample's diameter on the $B_0$ homogeneity.** [1]H FLASH MRI of 5 mm (a), 10 mm (b), and 16 mm (c) OD sample tubes filled with 20 mm of acetone and corresponding non-localized [1]H NMR spectra (d). The Lorentzian function fits (lines on d) yielded FWHM of 1.047, 1.574, and 2.492 ppm for 5, 10, and 16 mm sample tubes, respectively, under a $B_0$ magnetic field of 0.4454 T. The (0,0) coordinate corresponds to the isocenter of the magnet.

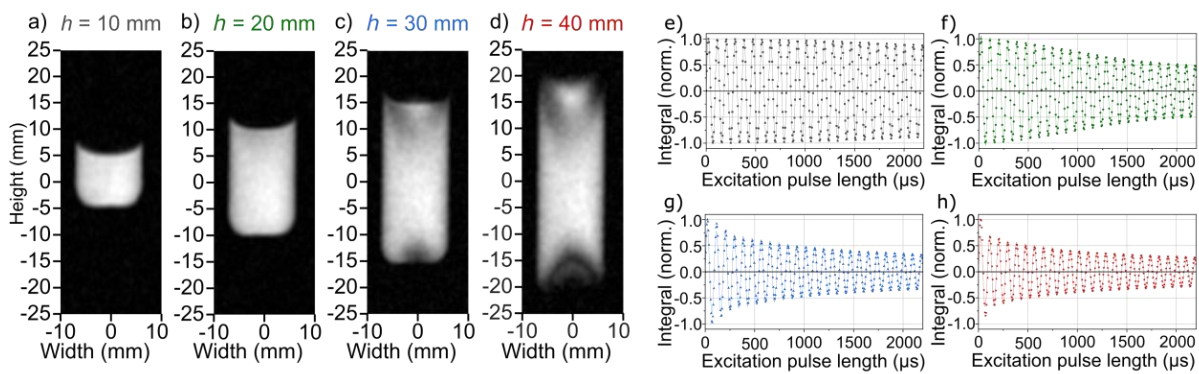

**Figure 3. Effect of the sample height on $B_0$ and $B_1$ homogeneity.** [1]H FLASH MRI (a-d) and nutation curves (e-h) of acetone in the 16 mm OD reactor filled to heights, $h$, 10 (a, e), 20 (b, f), 30 (c, g), and 40 mm (d, h). A damped wave function $Ae^{-kt}\sin(2\pi t_{pulse}/T)$ was fitted (continuous line) to the data and yielded $T$ = (90.30 ± 0.14) μs and $k$ = 57.8, 369.7, 521.8, and 531.9 s[-1]. $B_0$ and $B_1$ fields for samples up to $h$ of 20 mm and OD of 16 mm are sufficiently homogeneous: in the case of h = 20 mm, the signal decreased by 6.7% after 5 periods.

The Hahn equation[41], $M_T = M_0 e^{-R_2^{obs} 2n\tau}$, where $M_T$ is the transverse magnetization, $M_0$ is the initial magnetization, $R_2^{obs}$ is the observed $T_2$ relaxation decay rate, and $n$ is the number of refocusing pulses was used to fit the signal decay kinetics as a function of $n$ for each $\tau$.

In a perfectly stable and homogeneous static magnetic field, the signal decay shouldn't vary with echo interval for acetone; in our case, however, it did change. The $R_2^{obs}$ values were fitted (**Figure 4** - red curves) between 0-300 ms of $2\tau$ range with the equation accounting for $B_0$ inhomogeneity and diffusion[39] $R_2^{obs} = \frac{1}{T_2} + D^*(2\tau)^2$, where $D^*$ is the effective diffusion-homogeneity coefficient, which is given by $D^* = \frac{1}{12}\gamma^2 G^2 D$, where $\gamma$ is the gyromagnetic ratio, $G$ is the magnetic field gradient of the system caused by $B_0$ inhomogeneity, and $D$ is the ideal diffusion coefficient. When $D^*$ is compared with the previously assessed FWHM (**Figure 2d**) of the corresponding samples (and the origin point (0,0) as a point to pass through), the linear fit showed a high correlation with an $R^2$ = 0.99895 (**Figure 4**-inset). This supported the hypothesis of the observed $R_2^{obs}$ dependency.

For all ODs and sample heights of 20 mm, echo times below 50 ms appeared stable enough, providing similar observed $T_2$; hence, for polarization transfer, the echo times should not exceed this value. The characterizing properties of the large bore NMR system, such as FWHM, $T_2^*, T_2$, and $D^*$ with different OD tubes are summarized in **Table 1**.

**Table 1**. FWHM, $T_2^*$, $T_2$, and $D^*$ values measured for 5, 10, and 16 mm OD tubes filled with acetone to $h$ of 20 mm. Measurements were done with acetone at 40°C and $B_0$ = 0.4454 T. $T_2^*$ is estimated from FWHM as $1/(\pi \cdot \text{FWHM})$. $T_2(\tau = 0)$ and $D^*$ were obtained from CPMG measurements.

| Reactor diameter (mm) | FWHM (Hz) | FWHM (ppm) | $T_2^*$ (s) | $T_2$ ($\tau = 0$) (s) | $D^*$ (s[-1]) |
|---|---|---|---|---|---|
| 5 | 1.05 | 19.9 | 1.32 | 4.34±0.11 | 1.27±0.13 |
| 10 | 1.57 | 29.9 | 0.94 | 4.61±0.07 | 1.56±0.08 |
| 16 | 2.50 | 47.3 | 0.45 | 3.54±0.07 | 2.09±0.13 |

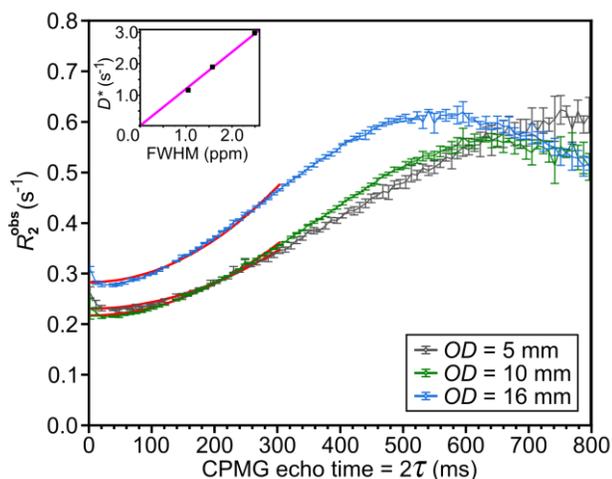

**Figure 4.** $T_2$ **relaxation of acetone.** Observed $T_2$ relaxation rate $R_2^{obs}$ as a function of CPMG echo time, $2\tau$, for 5, 10, and 16 mm OD tubes filled with acetone to $h$ of 20 mm. 169 echoes were measured for each $\tau$ value. $R_2^{obs}$ is a result of a mono-exponential decay fitting, $M_T = M_0 e^{-R_2^{obs} 2n\tau}$ to the obtained signal decay for each $\tau$. For $2\tau > 300$ ms, Hahn equation, $R_2^{obs} = \frac{1}{T_2} + D^*(2\tau)^2$, was fitted, yielding an effective diffusion-homogeneity coefficient $D^*$. The insert demonstrates the correlation between $D^*$ and FWHM of the samples (pink line - linear fit). Whiskers indicate the standard deviation of 5 repeated experimental measurements.

**Reactor design and simulation**

We designed 10 different reactors with variable properties to accommodate various experimental conditions and setups (**Figure 5**). The reactors were designed to increase the attainable pressure of pH$_2$, enhance hyperpolarization yield by accelerating the hydrogenation rate, increase volume, and reduce the costs associated with previously used, expensive proprietary high-pressure NMR tubes.

While the aim was to increase the hyperpolarized contrast agent volume by using 16 mm OD tubes, our designs also included 5 and 10 mm versions for comparison. The smaller tubes are compatible with widely used commercial NMR tubes and high-resolution NMR systems. All reactor designs with 4 ports (**Figure 5a**) were compatible with narrow bore (NB) NMR devices: the "spinnerless cap" can be used by itself; pressure caps should be used with adequately long tubes and corresponding alignment spinners.

The 4-port pressure caps were designed to allow for the injection of the precursor, infusion of purification components, pH$_2$ bubbling, and exhaust. Due to the physical constraints of the NB NMR devices and the limited amount of precursor in these reactors, it was not possible to implement a large vacuum port in these designs. Alternatively, the exhaust port can be used as a vacuum port for solvent evaporation.

The fifth port on large designs was explicitly included to increase the evaporation rate of the solvent. By increasing the diameter of the evaporation port, more solvent in gas form will be able to evaporate in a shorter time. This evaporation step is needed for the purification of the hyperpolarized solution, if the evaporation of the solvent is chosen as the purification method[27,28,43]. This expanded evaporation port is expected to shorten the time needed for this step.

While the glass tubes are cost-effective and commercially available, the property of the glass material makes them unusable under high pressure. To accommodate the weak mechanical properties of the glass material, the thickness of the walls must be increased, which in turn reduces the sample volume. Polymer tubes with a wall thickness of 1 mm, as opposed to 1.75 mm for glass tubes, offer a larger sample volume of ~12%. Also, the 10 mm OD standard glass tubes can't be used under pressures higher than 15 bar. In this case, polymer tubes are the only option for high-pressure hydrogenation, except sapphire tubes, which are expensive (10 mm OD, 7 mm ID, 7" length, ~750€/piece). Although polymers are not as chemically resistant as glass to solvents like acetone, polymer tubes are the only cost-effective solution.

The designs were created and simulated with Autodesk Inventor Professional 2024 (Autodesk Inc.). The pressure inside the reactors was applied as a pressure load onto the inner walls of the reactors and the pressure caps. According to our simulations, the maximum stress values under 100 bar of pressure for PEEK, PSU, and PEI reactors were 81.13, 91.61, and 93.21 MPa, respectively. The maximum displacement values were 637.2, 280.7, and 323.5 μm for 5, 10, and 16 mm tubes, respectively.

Out of 30 stress simulations under 100 bar of pressure load that were analyzed, only three simulations with PSU material resulted in values higher than the average yield tensile strength of the material[44], which are shown in **bold** in **Table 2**. As PSU material has a lower mechanical stress value than PEEK and PEI, and doesn't provide any extra advantages, the reactors were produced from PEEK and PEI materials.

**Injection reproducibility**

Before moving to hyperpolarization, we measured the reproducibility of the precursor injection into the reaction vessel with designed pressure caps. In this experiment, 5, 10, and 16 mm glass tubes were filled to a height of 20 mm with acetone, and subsequently, the signal with a 90° pulse was measured. After that, the sample was ejected by pressurizing the tube and leaving one outlet open through which liquids were ejected. This procedure was repeated automatically 10 times, and the entire experiment was repeated 3 times for each tube size (**Figure 6a** - shuttle). The standard deviations for each tube in the case of the injection series were 6.71%, 1.41%, and 0.63% for 5, 10, and 16 mm tubes, respectively.

The values of these integrals were also compared with the weighted amount of acetone (**Figure 6b**). The reference series was marginally more stable than the injection series, suggesting a slight increase in variability. Specifically, for the reference series, the standard deviation values were 2.50%, 2.23%, and 1.04% for 5, 10, and 16 mm tubes, respectively.

A high correlation between the obtained signal and the measured volume was observed. The measured volume (calculated from the weight) for 5 mm, 10 mm, and 16 mm tubes corresponded to the sample volumes of (0.265 ± 0.007) mL, (1.013 ± 0.023) mL, and (2.215 ± 0.023) mL, respectively.

Since the magnet exhibited a significant thermal drift of 1-2 ppm/mK, the resonance frequency adjustment was introduced right before excitation and measurement of the spectrum, which significantly increased long-term reproducibility.f

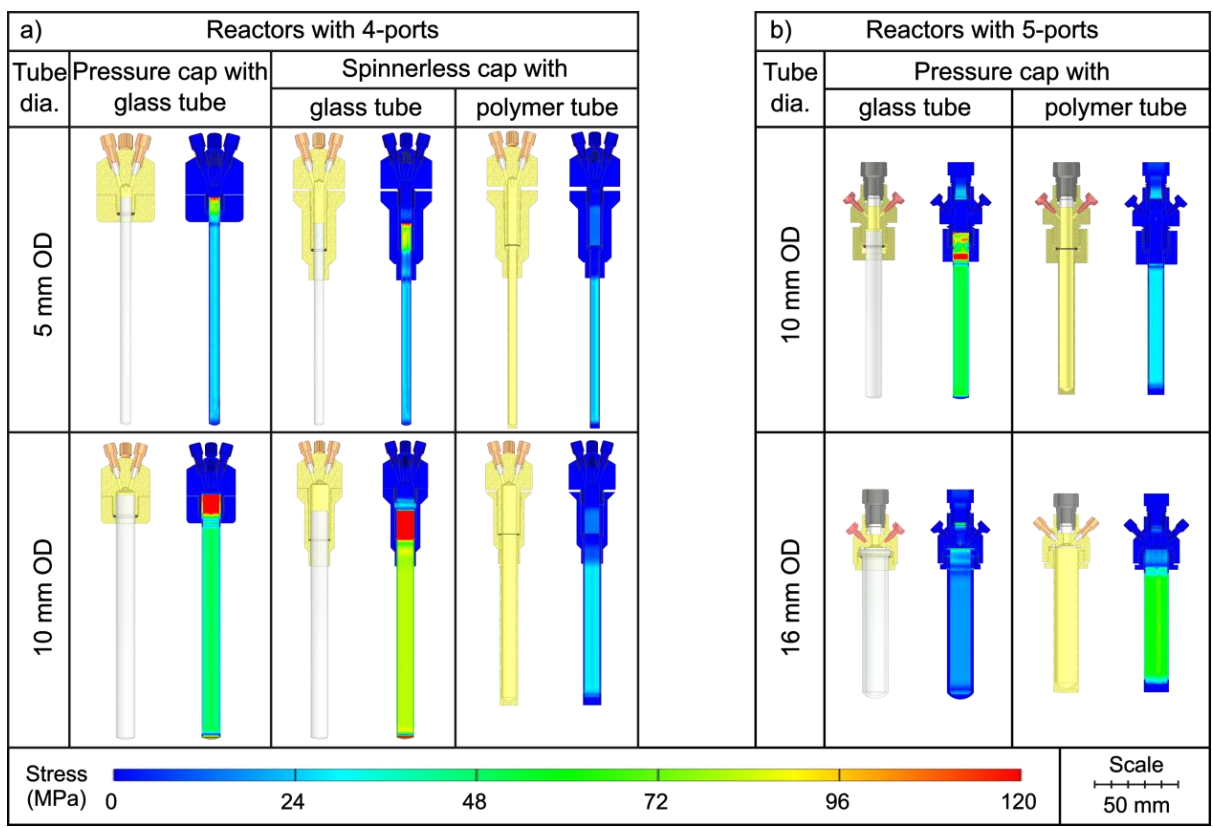

**Figure 5. Overview of the designed pH$_2$ reactors and pressure caps.** (a) Narrow bore 4-port NMR compatible pressure caps for 5 and 10 mm OD NMR tubes. When sufficiently short tubes are used, an additional spinner is not necessary, as it is integrated into the pressure cap (spinnerless cap). (b) 5-port pressure caps compatible with 10 mm and 16 mm OD tubes, featuring one extra-large port (through-hole diameter 4 mm) designed for applications such as solvent evaporation via vacuuming. The smaller 4 ports were designed for 1/16" or 1/32" tubing.

**Table 2.** Stress parameters of the designed pH$_2$ reactors and pressure caps. Maximum Von Mises Stress (MPa) and displacement (μm) values of the simulated reactors when 100 bar pressure is applied. (PC: pressure cap, SC: spinnerless cap, PT: polymer tube, see **Figure 5**). The average tensile strengths for PEEK (97.10 MPa) and PEI (114.00 MPa) were higher than simulated values. Still, for PSU (84.80 MPa), on some setups, the simulated values were higher than average yield tensile strengths (shown in bold), meaning that the structure was likely to fail. No simulated displacement value exceeded the elongation at yield, indicating that the elasticity of the structures is protected.

|  | OD (mm) | Tubes | Max Von Mises Stress (MPa) | | | Displacement (μm) | | |
| --- | --- | --- | --- | --- | --- | --- | --- | --- |
|  |  |  | PEEK | PSU | PEI | PEEK | PSU | PEI |
| 4 ports | 5 | PC | 36.50 | 43.7 | 38.85 | 92.7 | 25.1 | 31.9 |
|  |  | SC | 36.02 | 46.21 | 40.00 | 134.2 | 37.2 | 49.2 |
|  |  | PT | 52.06 | 52.34 | 52.16 | 428.0 | 159.3 | 155.7 |
|  | 10 | PC | 63.80 | 58.37 | 59.18 | 63.8 | 15.6 | 20.7 |
|  |  | SC | 81.06 | **91.61** | 82.57 | 283.7 | 64.5 | 87.8 |
|  |  | PT | 47.86 | 47.91 | 47.88 | 535.8 | 178.1 | 183.6 |
| 5 ports | 10 | PC | 68.57 | **89.45** | 80.54 | 364.3 | 119.5 | 154.2 |
|  |  | PT | 77.38 | **89.30** | 80.08 | 637.2 | 280.7 | 323.5 |
|  | 16 | PC | 39.63 | 64.83 | 51.07 | 118.5 | 51.2 | 60.9 |
|  |  | PT | 81.13 | 82.22 | 93.21 | 571.1 | 209.6 | 199.6 |
| Max. average yield values from literature[44] | | | 97.10 | 84.80 | 114.00 | 6585 | 3520 | 8075 |

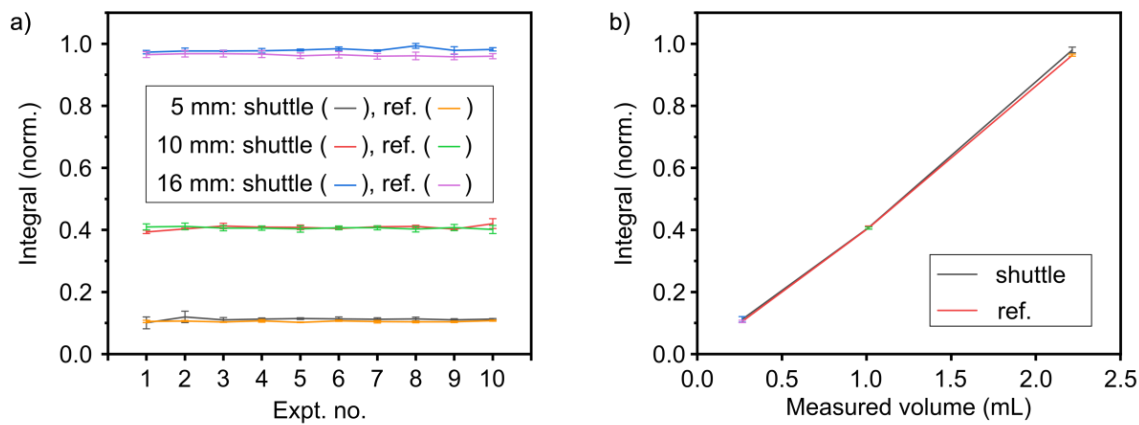

**Figure 6. Sample injection reproducibility**. (a) Reproducibility of signal integrals from 90° pulse experiments for reference (ref., manually pipeted volume) and automatically shuttled (shuttle) samples for 5, 10, and 16 mm tubes. Measurements show minimal variation across repetitions (with and without liquid shuttling). (b) The average integral of the spectra and standard deviation of all 30 spectra for each tube were plotted against the sample volume calculated from the measured sample weight. A correlation of 0.99994 was obtained.

To assess the stability of the system, we repeated the experiment with manual pipetting of acetone into the same tubes and then measured the sample 10 times. This process was repeated three times for each tube (**Figure 6** -ref.).

**Optimizing ¹H spin order transfer**

As a model reaction, we chose the hydrogenation of vinyl acetate, which yields ethyl acetate (EA) as the product. Fully protonated (VA-$h_6$) and deuterated (VA-$d_6$) precursors were compared.

Out-of-phase echo SOT with three composite refocusing pulses ($45_X$-[$\tau$-$90_X 180_Y 90_X$-$\tau$]$_{n=3}$-FID) was used (**Figure 7a**). This sequence converted the antiphase PASADENA spectrum into an in-phase one, which was especially useful for inhomogeneous magnetic fields[39,45–50]. All experiments presented here and in the following sections were performed using a 16 mm tube and pressure cap reactor made of PEEK material.

Using the same injection protocol as developed and described previously, we again performed three series of 10 experiments for VA-$h_6$ and VA-$d_6$, where $\tau$ varied (**Figure 7c, d**). The optimal $\tau$ of 6.9 ms and 10.6 ms for EA-$h_6$ and EA-$d_6$ were found, and the obtained kinetics were well fitted with the simulations.

Finally, using VA and optimized $\tau$, the polarization experiment was again performed in 2 series, 10 experiments in each: resulting relative standard deviation of the mean was 6.79% (**Figure 7b**).

**Temperature calibration**

To determine the actual temperature of the sample, a thermocouple was used in conjunction with a data logger (Voltcraft K204 Thermometer, Conrad Electronic SE), which was passed through an unused port and fixed with a fitting. Then the injection cycles were repeated 3 times for different temperature settings of the heater (**Figure 8a**). It is not possible to heat the sample higher than the boiling point of acetone (56.2°C, **Figure 8a**; grey dashed line) when the exhaust of the system is open, i.e., at ambient pressure. When, however, the exhaust was closed (**Figure 8b**-80°C*), the pressure in the reactor could go higher, which allowed the temperature in the reactor to reach up to 63°C in 100 seconds.

It can be seen that the system doesn't reach the boiling point for any heater settings within 40 seconds; hence, all further experiments were conducted with this heating time (**Figure 8a**).

**Temperature effect on PHIP**

Hydrogenation times from 2 to 60 s were also investigated with varying temperatures (**Figure 8b**). With higher temperatures, it has been observed that shorter hydrogenation times are required to achieve the maximal signal.

**Pressure and p$H_2$ flow effects on reaction kinetics**

Then we assed the effect of the p$H_2$ pressure and flow on the PHIP signal (**Figure 9**). p$H_2$ pressure had the advantage of increasing the amount of dissolved p$H_2$ in the acetone. This had the potential advantage of increasing the interaction of p$H_2$ and the catalyst, leading to accelerated hydrogenation. While the p$H_2$ flow rate had a direct effect on the amount of delivered p$H_2$ gas, it also affected the bubble size, hence contributing to the replenishment of p$H_2$ in the solution.

It can be seen that the increased flow rate increased polarization only in the case of the 40°C system: the signal grew by 37.7% when the flow changed from 1 L/min to 2.5 L/min. With the 80°C system, the flow rate didn't have any obvious effect. The optimized parameters (92% p$H_2$ conversion, 20 bar p$H_2$ pressure, 2 L/min flow, 4 s $t_b$, 55°C sample temperature for 2.2 mL sample in 16 mm OD microwave tube, 50 mM VA conc., 5mM Rh conc.) allowed us to reach ¹H hyperpolarization of EA-$d_6$ up to P = 31.3%.

**Discussion**

In summary, the successful design and implementation of an automated, larger-volume hyperpolarizer capable of producing high polarization levels for biologically relevant substances while operating at high pressure were demonstrated.

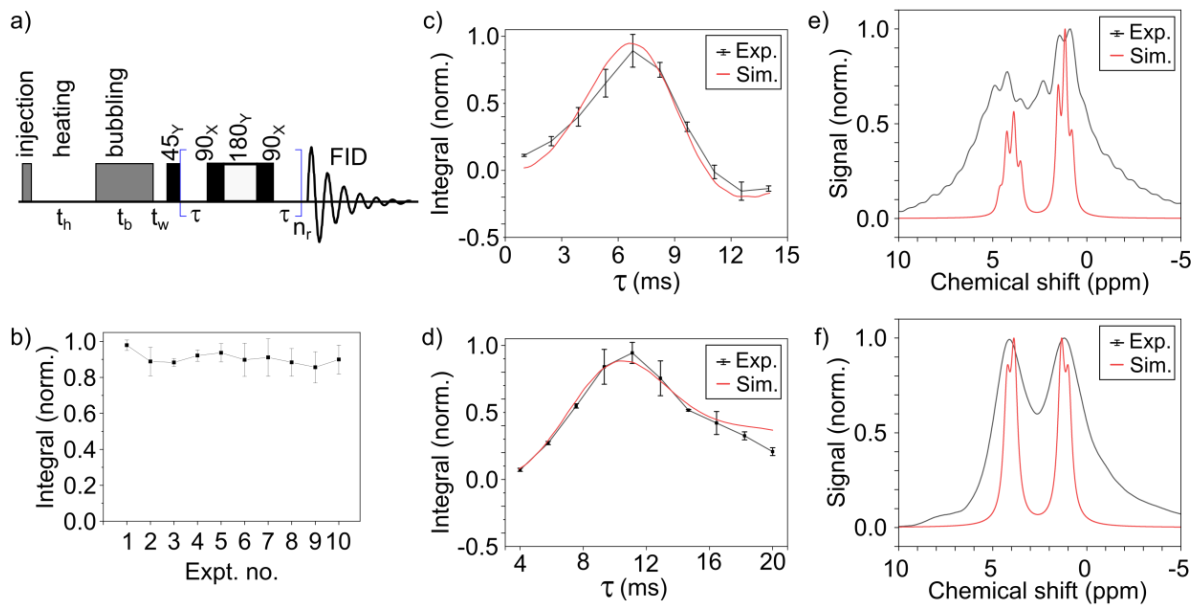

**Figure 7. PHIP of VA with out-of-phase echo SOT.** Experiment scheme (a), hyperpolarization repeatability (b), and integrals of SOT($\tau$) dependences for VA-$h_6$ (c) and VA-$d_6$ (d) and corresponding spectra (e and f). a) The experiment started with the injection of the precursor, a waiting period for heating of the sample, $t_h$, the hydrogenation (bubbling) of the sample during the bubbling time, $t_b$, a waiting period for the disturbed solution to settle down, $t_w$, and application of the OPE sequence with $n_r$ = 3 refocusings and acquisition of the signal. b) The reproducibility of the hyperpolarization process is measured by conducting two series of 10 experiments each consisting of injecting, heating, bubbling, SOT, FID, and ejecting. The resulting standard variance for the obtained integrals was 6.79%. c,d) Simulated (red lines) and experimental signal of EA-$h_6$ (c) and EA-$d_6$ (d), and corresponding spectra at optimum $\tau$ (e, f). The homogeniously broadened linewidth in simulated spectra was set to 18.717 MHz. Here, $t_b$ was 10 s, $t_h$ was 80 s, $t_w$ was 0.1 s, and the temperature of the heater was set to 40°C.

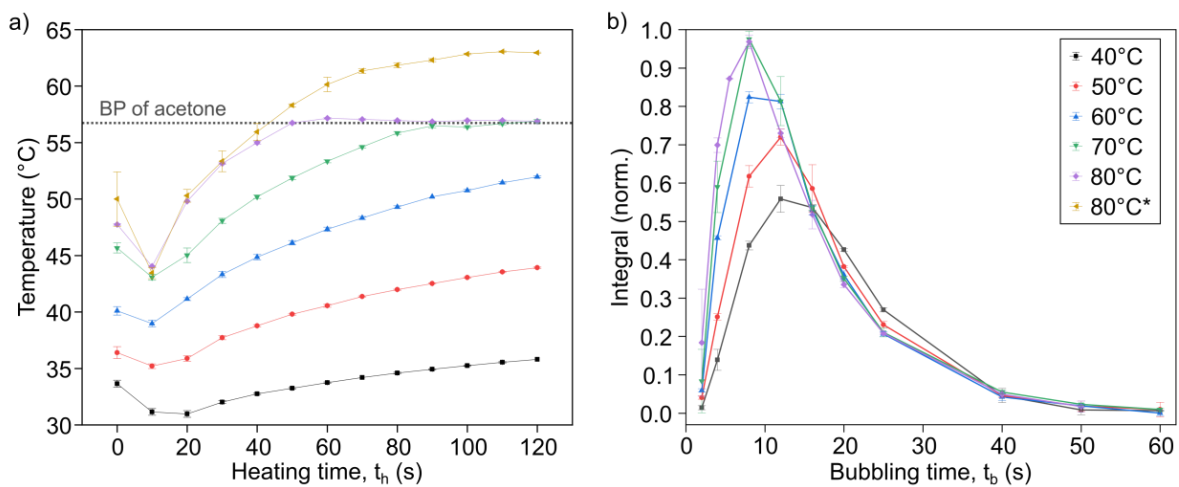

**Figure 8. Temperature calibration and its effect on PHIP.** a) Temperature change recorded after injection of the 2.2 mL sample in a 16 mm OD tube under various heating temperature settings. The 80°C* temperature denotes that the exhaust of the reactor is closed, so pressure can build up inside the reactor, which allows the acetone to reach higher temperatures. b) Signal yield after hydrogenation of VA-$h_6$ and application of optimized out-of-echo SOT after variable $t_b$. The sample consists of 50 mM VA, 5 mM Rh, and 2.2 mL of acetone. It can be observed that as the temperature increases, shorter bubbling times become more efficient.

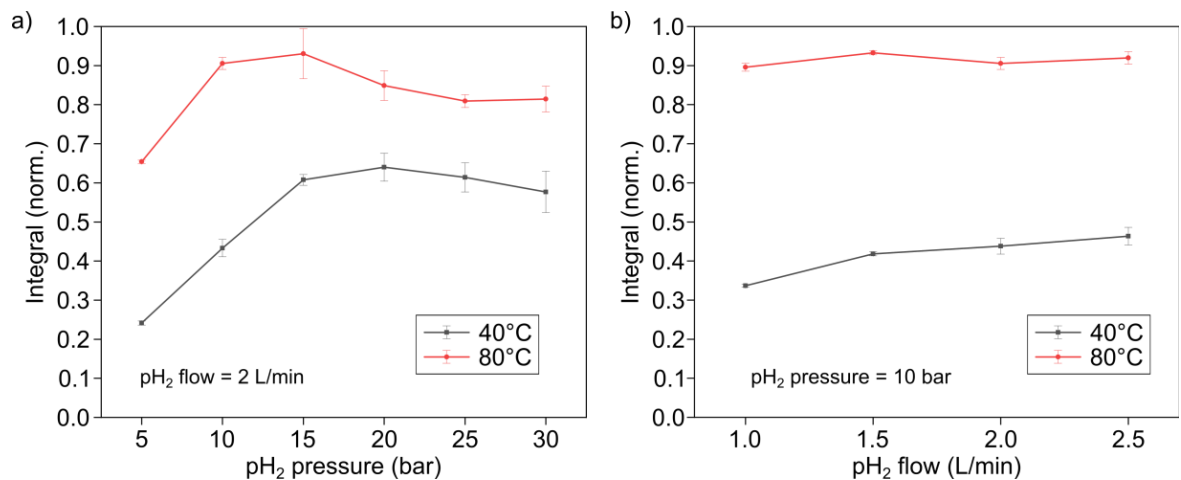

**Figure 9. Effect of hydrogenation pressure (a) and pH$_2$ flow rate (b).** Bubbling time, t$_b$, was 12 s for 40°C heating temperature, temperature at the time of bubbling was 32.8 °C, and 8 s for 80°C heating temperature, temperature at the time of bubbling was 55.0°C(see Figure 8b for optimum t$_b$). For the hydrogenation pressure effect, the flow of pH$_2$ was 2 L/min. For the pH$_2$ flow rate effect, the hydrogenation pressure was 10 bar. For all experiments, VA concentration was 50 mM, and the catalyst concentration was 5 mM.

The system's unique integration of automated injection and gas handling, sample heating, high pressure and high volume availability, and rapid sample transfer allowed us to reach polarization levels of up to 31.3% for EA-$d_6$ in acetone in a duty cycle of 80 s with an injection automation error of ~14 μL (0.63%) and a standard deviation of polarization reproducibility of 6.79% from the mean.

The used pulse sequence yields 50% polarization in the ideal case and 100% of pH$_2$, while 44.6% for 92% pH$_2$, which were used. This demonstrates that we achieved 70% of the theoretical maximum (31.3/44.6 = 70%) while also increasing the polarization volumes to 2.2 mL, which is sufficient for preclinical use. The achieved $^1$H molar polarization[51] is calculated to be 31.3 mM for 2.215 mL of solution. If another sequence is used, which focuses polarization on one proton, then the double polarization could be produced[52], which corresponds to 62.6% $^1$H in-phase polarization; however, the molar polarization will stay the same.

The $B_0$ and $B_1$ magnetic field homogeneity analysis results showed that although the FWHM of the 16 mm reactor was 2.492 ppm, the acquired images were of relatively good quality. Larger volumes of samples (h ≥ 30 mm) yielded noticeably worse image quality, as well as faster decaying nutation curves, compared to samples with h ≤ 20 mm, due to the poor homogeneity of $B_0$ and $B_1$.

The automation allowed us to optimize the following parameters reproducibly: sample temperature, pH$_2$ pressure, pH$_2$ flow, hydrogenation time, and SOT interpulse delays. The maximum was achieved at 15 bar of pH$_2$ and its flow of 1.5 L/min, 80°C heater temperature (or 55.0°C sample temperature), and a hydrogenation time of 8 s. Since the increase of pH$_2$ pressure and flow did not improve polarization yield, we concluded that the 5 mM of used catalyst is already saturated with the available amount of pH$_2$. The estimated concentration of pH$_2$ in the sample under 15 bar of hydrogenation pressure and 55°C of temperature is ~80 mmol/L[53]. The decline in polarization yield when the pressure exceeds 15 bar (**Figure 9a**) is tentatively attributed to vigorous bubbling and displacement of the sample mixture out of the volume of interest (VOI), resulting in less signal being acquired.

Designed reactors with multiple connectors, including one large port for vacuuming, enabled robust injection, pressurization, evaporation, and purification of the samples under relatively high pressure. The 16 mm microwave tube was found to be convenient as a reactor and novel solution proposed here, as it provided excellent chemically stable properties and had a low cost of approximately 2€ per tube, which was similar to economy 5 mm NMR tubes but permitted experiments with pressures of up to 41 bar (600 psi) and at larger volumes.

Future studies will focus on automated, fast, and reliable purification of the hyperpolarized contrast agent following methodologies developed previously[28,37,40,54].

## Methods

### Chemicals

As the catalyst of the hyperpolarization process, [1,4-Bis-(diphenylphosphino)-butan]-(1,5-cyclooctadien)-rhodium(I)-tetrafluoroborat ([Rh], 341134, Sigma-Aldrich) was used. Vinyl acetate (VA, 48486, Merck) and vinyl acetate-$d_6$ (VA-$d_6$, D-4097, CDN Isotopes) were used as precursors. Acetone (320110, Sigma-Aldrich) and acetone-$d_6$ (444863, Sigma-Aldrich) were used as solvents for the preparation of the samples. The samples were prepared by mixing 50 mM of one of two precursors (VA, VA-$d_6$) with 5 mM [Rh] in acetone or acetone-$d_6$. The prepared sample was loaded into a 20 mL syringe and used in the hyperpolarization experiment.

The pH$_2$ that was used for hydrogenation was prepared using an in-lab built pH$_2$ generator that used liquid nitrogen as coolant (52% enrichment)[20] or with a two-stage cryosystem-based generator (92% enrichment)[22].

## Hydrogenation

**Bubbling time:** The bubbling time is related to the hydrogen saturation of the solution, which in turn affects the hyperpolarized precursor amount. Longer bubbling times usually mean higher amount of precursor to be hydrogenated, but also give more time for hyperpolarized product to relax. In this study, 10 different bubbling times (2, 4, 8, 12, 16, 20, 25, 40, 50, 60 s) were investigated. For 80°C temperature experiments, the 5.5 s bubbling time was added, and the 60 s bubbling time was omitted, while increased temperatures shorten the exchange time.

**pH$_2$ pressure:** Standard NMR tubes can be used up to 10 bar of pressure, but for higher pressures, a custom NMR tube out of PEEK and PEI with flanges had to be produced. In order to find the optimal hydrogenation pressure, 5, 10, 15, 20, 25, and 30 bar of pH$_2$ hydrogenation pressure values were tested.

**pH$_2$ flow:** The flow rates of 1.0, 1.5, 2.0, and 2.5 L/min were used. The flow of the pH$_2$ was measured with the flow controller (C100L, Sierra Instruments Inc.).

**Sample temperature:** The temperature of the reaction has a shortening effect on the hydrogenation kinetics[37], which can be calculated as[39]:

$$[PCH] = \frac{[PC_0]P_0}{1 - R/k_1}(e^{-\tau_b R} - e^{\tau_b k_1})$$

where [PCH] is the precursor hydrogenation concentration, [PC$_0$] is the initial precursor concentration, P$_0$ is the reached polarization, R is the reaction rate, k$_1$ is the first-order rate constant of the reaction, and $\tau_b$ is the pH$_2$ bubbling time. It is also proven that elevated temperatures have a positive effect on polarization levels[30].

In order to elevate the temperature of the samples, a sample heating system (WTHA 1 - Weller Tools GmbH) with hot air was implemented. The samples were heated up to 40°C, 50°C, 60°C, 70°C, and finally 80°C, which was the temperature limit of the sample heating system.

## Reactor design, simulation, and production

**General description of the hyperpolarization reactor.** We experimented with 4-port reactors, which can enable various liquid sample shuttling and delivery of pH$_2$, and with 5-port reactors, where an additional large orifice tubing can be connected for rapid vacuuming of the solvent. All polymer parts were machined at the Central Workshop of the Biology Department and the Central Workshop of the Physics Department, CAU, according to the technical drawings (Supplementary Information). The constituents of such reactors were described below.

**Glass reaction vessel.** 5 mm (Boroeco-5-7, Deutero GmbH), 10 mm (Boroeco-10-7, Deutero GmbH), and 16 mm (908035, CEM GmbH) OD glass tubes were convenient vessels for hydrogenation experiments due to their chemical resistivity, low costs of ~1-2€ per tube, and large opening (4.2 mm ID for 5 mm OD, 9.1 mm ID for 10 mm, and 12.5 mm ID for 16 mm OD) compared to high-pressure tubes e.g., high pressure 10 mm NMR tube has an opening neck of 0.8 mm inner diameter (513-7PVH-7, Wilmad).

The designed pressure caps and reactors were suitable for standard 5 mm NMR tubes or 10 mm NMR tubes; the caps with integrated spinners were suitable for 4-inch 5 mm NMR tubes (Boro-5-103.5-o, Deutero GmbH). 4-inch 10 mm NMR tubes were uncommon, and most of the time, they had to be custom-cut by the supplier. The 16 mm microwave reactor tubes used with pressure caps were pressure-rated up to 600 psi (41 bar). The pressure caps can also be modified to suit, e.g., 35 mL microwave tubes with a 30 mm OD to enable even larger sample volumes (pressure-rated up to 20 bar, 24.5 mm ID, 909036, CEM GmbH).

**4-port pressure caps:** Two polymers, PEEK (polyetheretherketone) and PEI (polyetherimide), were used for pressure caps and reactor vessels because of their high tensile strengths and chemical resistance to the solvents. 30 mm diameter PEEK and PEI polymer rods (L. Buck & Sohn GmbH & Co.) were machined following the technical drawings (**SI, Figure S2-S6**). These pressure caps were equipped with four PEEK fittings and ferrules for 1/16" tubing (F-333, F-142, IDEX Health & Science, LLC). Such a configuration was designed for 5 and 10-mm OD NMR tubes. The assembled pressure cells were compatible with the standard bore Bruker system. Optionally, one can use a PEEK sleeve (F-233, IDEX Health & Science, LLC) to decrease the diameter of the fitting tubes to 1/32". Because of the integration of the spinner geometry into the pressure cap design, the use of a spinner can be eliminated with a 4-inch-long glass NMR tube.

**5-port pressure caps (4-ports plus vacuuming port):** To enable rapid vacuuming, a larger port was necessary. This was designed only for 10 and 16-mm OD tubes. The current configuration of the pressure cap had a maximum diameter of 29.8 mm, which was incompatible with the standard bore Bruker system. However, it fits a wide-bore Bruker system and is compatible with our low-field MRI system. The vacuum port was introduced in the middle of the cells with a PEEK nut and ETFE ferrule (U-655, U-650, IDEX Health & Science, LLC, max 17 bar), enabling a connection to a vacuum pump through ¼" inch PA tubing (PA 1/4 SCHWARZ, Landefeld Druckluft und Hydraulik GmbH). When the system should withstand higher than 17 bar pressures, one can instead use a brass nut (GT 144 MS, Landefeld Druckluft und Hydraulik GmbH) to connect ¼" inch PA tubing to the vacuum port.

**Polymer reaction vessel:** To increase hydrogenation pressure, polymer tubes are designed and produced from PEEK and PEI polymers. The simulations also include PSU (polysulfone) material, but the relatively weak material properties and low resistance to some solvents, e.g. acetone, precluded us from using it as a reaction vessel material. 5 mm polymer reaction vessels are also not produced, due to their low volume and relatively hard machining processes originating from their small diameters and high tube lengths.

**Sealing:** To enable gas sealing of straight 5 and 10 mm OD NMR tubes, we used O-rings made from FFKM polymer for universal solvent resistance: 4.5×1.5 mm (1087924, Alwin Höfert KG) and 9.25×1.78 mm (1008515, NH O-RING GmbH & Co. KG), respectively. For the 16 mm OD tube with a flange, we used a PTFE (polytetrafluoroethylene) flat O-ring 13.5×2.25 mm dimensions with 2 mm height (DR 14 TE, Landefeld Druckluft und Hydraulik GmbH). Because of the geometry difference and the way the tubes were connected to the pressure cap, in the case of straight 5 and 10-mm flangeless NMR tubes, when the O-rings get wet, the tube can slide from the pressure cap. This was not the case for the 16 mm OD tube with the flange, in which case, only sealing can be compromised when the O-ring is wet.

**Experimental reactor pressure ranking:** While the standard glass NMR tubes can withstand 10 bar of pressure (not specified by manufacturer), the system with the 16 mm OD microwave tube can withstand at least 34 bar of pressure (specified by manufacturer), which is the maximum pressure rating of the valves (P-782, IDEX Health & Science, LLC) used in the injection of liquids. The systems with 5 mm and 10 mm glass NMR tubes were pressurized up to 10.1 bar, and the system with the 16 mm microwave tube was pressurized up to 30.4 bar. No noticeable problem was seen.

**Simulations of the designed reactors:** The pressure caps and reactors were designed in Autodesk Inventor Professional 2024 (Autodesk Inc.), and the designs were loaded with 100 bar of pressure load on the related surfaces. Although this pressure value was much higher than the intended pressure value during experiments, the safety factor was kept higher to accommodate the errors of production, machining, material fatigue after pressurizing and depressurizing several cycles, and solvent degradation of the polymers, which could not be simulated in finite element method (FEM) analysis. The mesh of the FEM analysis was configured to have a 5% average element size and a 2.5% minimum element size of the models.

**Simulation of spectra and OPE performance:** The expected spectra for EA-$h_6$ and EA-$d_6$ were simulated and compared with the experimental results (**Figure 7e-f**). The spectra were homogenously broadened to imitate the experimental conditions. The OPE performance was also simulated and compared with experimental results. An exponential decay function, depending on $\tau$, was applied to the magnetization from the simulations to suppress oscillations.

**Quantification of hyperpolarization:** The polarization level was calculated using the reference thermal signal. For the thermal signal, the sample was measured 10 minutes after the hyperpolarization process, ensuring that all of the hyperpolarization signal had relaxed. After this measurement, the same sample was analyzed with a high-resolution, 9.4 T NMR system with a 5 mm broadband probe (400 MHz WB, NEO; BBFO, Bruker) to determine the amount of EA-$d_6$ in the sample. The signal ratio of EA-$d_6$ to the whole spectrum (**SI, Figure S7**) was used as a correction coefficient between the thermal and hyperpolarized signal measured in the polarizer in situ. This is necessary because portable MRI does not provide sufficient spectral resolution to assess the signal from EA-$d_6$ exclusively.

## ASSOCIATED CONTENT

**Supporting Information**. The $B_1$ field distribution simulations, technical drawings and high-resolution $^1$H NMR spectra of the hydrogenated product is included in SI (.pdf). All used Matlab scripts to simulate OPE kinetics, NMR spectra and $B_1$ field distribution are included (.zip). This material is available free of charge via the Internet at http://pubs.acs.org.

All the raw data and technical drawings are available on Zenodo repository (https://doi.org/10.5281/zenodo.15807226).

## AUTHOR INFORMATION

### Corresponding Author

* andrey.pravdivtsev@rad.uni-kiel.de, jan.hoevener@rad.uni-kiel.de

### Author Contributions

ANP, YG: conceptualization, investigation, writing – original draft; YG: analysis; ANP, JBH, YG: writing, final versions; ANP and JBH: supervision, funding acquisition. All authors contributed to discussions and interpretation of the results and have approved the final version of the manuscript.

### Funding Sources

We acknowledge support from the German Federal Ministry of Education and Research (03WIR6208A hyperquant), DFG (555951950, 527469039, 469366436, HO-4602/2-2, HO-4602/3, HO-4602/4, EXC2167, FOR5042, TRR287). MOIN CC was founded by a grant from the European Regional Development Fund (ERDF) and the Zukunftsprogramm Wirtschaft of Schleswig-Holstein (Project no. 122-09-053).

## REFERENCES

(1) Khalil, A.; Kashif, M. Nuclear Magnetic Resonance Spectroscopy for Quantitative Analysis: A Review for Its Application in the Chemical, Pharmaceutical and Medicinal Domains. *Crit. Rev. Anal. Chem.* **2023**, *53* (5), 997–1011. https://doi.org/10.1080/10408347.2021.2000359.

(2) Haase, A.; Frahm, J.; Matthaei, D.; Hänicke, W.; Merboldt, K.-D. FLASH Imaging: Rapid NMR Imaging Using Low Flip-Angle Pulses. *J. Magn. Reson.* **2011**, *213* (2), 533–541. https://doi.org/10.1016/j.jmr.2011.09.021.

(3) Eills, J.; Budker, D.; Cavagnero, S.; Chekmenev, E. Y.; Elliott, S. J.; Jannin, S.; Lesage, A.; Matysik, J.; Meersmann, T.; Prisner, T.; Reimer, J. A.; Yang, H.; Koptyug, I. V. Spin Hyperpolarization in Modern Magnetic Resonance. *Chem. Rev.* **2023**, *123* (4), 1417–1551. https://doi.org/10.1021/acs.chemrev.2c00534.

(4) Hövener, J.; Pravdivtsev, A. N.; Kidd, B.; Bowers, C. R.; Glöggler, S.; Kovtunov, K. V.; Plaumann, M.; Katz-Brull, R.; Buckenmaier, K.; Jerschow, A.; Reineri, F.; Theis, T.; Shchepin, R. V.; Wagner, S.; Bhattacharya, P.; Zacharias, N. M.; Chekmenev, E. Y. Parahydrogen-Based Hyperpolarization for Biomedicine. *Angew. Chem. Int. Ed.* **2018**, *57* (35), 11140–11162. https://doi.org/10.1002/anie.201711842.

(5) Ardenkjær-Larsen, J. H.; Fridlund, B.; Gram, A.; Hansson, G.; Hansson, L.; Lerche, M. H.; Servin, R.; Thaning, M.; Golman, K. Increase in Signal-to-Noise Ratio of > 10,000 Times in Liquid-State NMR. *Proc. Natl. Acad. Sci. U.S.A.* **2003**, *100* (18), 10158–10163. https://doi.org/10.1073/pnas.1733835100.

(6) Ferrari, A.; Peters, J.; Anikeeva, M.; Pravdivtsev, A.; Ellermann, F.; Them, K.; Will, O.; Peschke, E.; Yoshihara, H.; Jansen, O.; Hövener, J.-B. Performance and Reproducibility of 13C and 15N Hyperpolarization Using a Cryogen-Free DNP Polarizer. *Sci. Rep.* **2022**, *12* (1), 11694. https://doi.org/10.1038/s41598-022-15380-7.

(7) Green, R. A.; Adams, R. W.; Duckett, S. B.; Mewis, R. E.; Williamson, D. C.; Green, G. G. R. The Theory and Practice

of Hyperpolarization in Magnetic Resonance Using Parahydrogen. *Prog. Nucl. Magn. Reson. Spectrosc.* **2012**, *67*, 1–48. https://doi.org/10.1016/j.pnmrs.2012.03.001.

(8) Schmidt, A. B.; Berner, S.; Braig, M.; Zimmermann, M.; Hennig, J.; Von Elverfeldt, D.; Hövener, J.-B. In Vivo 13C-MRI Using SAMBADENA. *PLoS ONE* **2018**, *13* (7), e0200141. https://doi.org/10.1371/journal.pone.0200141.

(9) Pravdivtsev, A. N.; Tickner, B. J.; Glöggler, S.; Hövener, J.-B.; Buntkowsky, G.; Duckett, S. B.; Bowers, C. R.; Zhivonitko, V. V. Unconventional Parahydrogen-Induced Hyperpolarization Effects in Chemistry and Catalysis: From Photoreactions to Enzymes. *ACS Catal.* **2025**, *15* (8), 6386–6409. https://doi.org/10.1021/acscatal.4c07870.

(10) Adams, R. W.; Aguilar, J. A.; Atkinson, K. D.; Cowley, M. J.; Elliott, P. I. P.; Duckett, S. B.; Green, G. G. R.; Khazal, I. G.; López-Serrano, J.; Williamson, D. C. Reversible Interactions with Para-Hydrogen Enhance NMR Sensitivity by Polarization Transfer. *Science* **2009**, *323* (5922), 1708–1711. https://doi.org/10.1126/science.1168877.

(11) Cunningham, C. H.; Lau, J. Y. C.; Chen, A. P.; Geraghty, B. J.; Perks, W. J.; Roifman, I.; Wright, G. A.; Connelly, K. A. Hyperpolarized $^{13}$C Metabolic MRI of the Human Heart: Initial Experience. *Circ. Res.* **2016**, *119* (11), 1177–1182. https://doi.org/10.1161/CIRCRESAHA.116.309769.

(12) Peters, J. P.; Assaf, C.; Mohamad, F. H.; Beitz, E.; Tiwari, S.; Aden, K.; Hövener, J.-B.; Pravdivtsev, A. N. Yeast Solutions and Hyperpolarization Enable Real-Time Observation of Metabolized Substrates Even at Natural Abundance. *Anal. Chem.* **2024**, *96* (43), 17135–17144. https://doi.org/10.1021/acs.analchem.4c02419.

(13) Sapir, G.; Steinberg, D. J.; Aqeilan, R. I.; Katz-Brull, R. Real-Time Non-Invasive and Direct Determination of Lactate Dehydrogenase Activity in Cerebral Organoids—A New Method to Characterize the Metabolism of Brain Organoids? *Pharmaceuticals* **2021**, *14* (9), 878. https://doi.org/10.3390/ph14090878.

(14) Gallagher, F. A.; Kettunen, M. I.; Day, S. E.; Lerche, M.; Brindle, K. M. $^{13}$C MR Spectroscopy Measurements of Glutaminase Activity in Human Hepatocellular Carcinoma Cells Using Hyperpolarized$^{13}$C-labeled Glutamine. *Magn. Reson. Med.* **2008**, *60* (2), 253–257. https://doi.org/10.1002/mrm.21650.

(15) Gallagher, F. A.; Woitek, R.; McLean, M. A.; Gill, A. B.; Manzano Garcia, R.; Provenzano, E.; Riemer, F.; Kaggie, J.; Chhabra, A.; Ursprung, S.; Grist, J. T.; Daniels, C. J.; Zaccagna, F.; Laurent, M.-C.; Locke, M.; Hilborne, S.; Frary, A.; Torheim, T.; Boursnell, C.; Schiller, A.; Patterson, I.; Slough, R.; Carmo, B.; Kane, J.; Biggs, H.; Harrison, E.; Deen, S. S.; Patterson, A.; Lanz, T.; Kingsbury, Z.; Ross, M.; Basu, B.; Baird, R.; Lomas, D. J.; Sala, E.; Wason, J.; Rueda, O. M.; Chin, S.-F.; Wilkinson, I. B.; Graves, M. J.; Abraham, J. E.; Gilbert, F. J.; Caldas, C.; Brindle, K. M. Imaging Breast Cancer Using Hyperpolarized Carbon-13 MRI. *Proc. Natl. Acad. Sci. U.S.A.* **2020**, *117* (4), 2092–2098. https://doi.org/10.1073/pnas.1913841117.

(16) Bowers, C. R.; Weitekamp, D. P. Transformation of Symmetrization Order to Nuclear-Spin Magnetization by Chemical Reaction and Nuclear Magnetic Resonance. *Phys. Rev. Lett.* **1986**, *57* (21), 2645–2648. https://doi.org/10.1103/PhysRevLett.57.2645.

(17) Bowers, C. R.; Weitekamp, D. P. Parahydrogen and Synthesis Allow Dramatically Enhanced Nuclear Alignment. *J. Am. Chem. Soc.* **1987**, *109* (18), 5541–5542. https://doi.org/10.1021/ja00252a049.

(18) Eisenschmid, T. C.; Kirss, R. U.; Deutsch, P. P.; Hommeltoft, S. I.; Eisenberg, R.; Bargon, J.; Lawler, R. G.; Balch, A. L. Para Hydrogen Induced Polarization in Hydrogenation Reactions. *J. Am. Chem. Soc.* **1987**, *109* (26), 8089–8091. https://doi.org/10.1021/ja00260a026.

(19) Jeong, K.; Min, S.; Chae, H.; Namgoong, S. K. Detecting Low Concentrations of Unsaturated C—C Bonds by Parahydrogen-induced Polarization Using an Efficient Home-built Parahydrogen Generator. *Magn. Reson. Chem.* **2018**, *56* (11), 1089–1093. https://doi.org/10.1002/mrc.4756.

(20) Ellermann, F.; Pravdivtsev, A.; Hövener, J.-B. Open-Source, Partially 3D-Printed, High-Pressure (50-Bar) Liquid-Nitrogen-Cooled Parahydrogen Generator. *Magn. Reson.* **2021**, *2* (1), 49–62. https://doi.org/10.5194/mr-2-49-2021.

(21) Du, Y.; Zhou, R.; Ferrer, M.-J.; Chen, M.; Graham, J.; Malphurs, B.; Labbe, G.; Huang, W.; Bowers, C. R. An Inexpensive Apparatus for up to 97% Continuous-Flow Parahydrogen Enrichment Using Liquid Helium. *J. Magn. Reson.* **2020**, *321*, 106869. https://doi.org/10.1016/j.jmr.2020.106869.

(22) Hövener, J.; Bär, S.; Leupold, J.; Jenne, K.; Leibfritz, D.; Hennig, J.; Duckett, S. B.; Von Elverfeldt, D. A Continuous-flow, High-throughput, High-pressure Parahydrogen Converter for Hyperpolarization in a Clinical Setting. *NMR Biomed* **2013**, *26* (2), 124–131. https://doi.org/10.1002/nbm.2827.

(23) Reineri, F.; Boi, T.; Aime, S. ParaHydrogen Induced Polarization of 13C Carboxylate Resonance in Acetate and Pyruvate. *Nat. Commun.* **2015**, *6* (1), 5858. https://doi.org/10.1038/ncomms6858.

(24) Reineri, F.; Cavallari, E.; Carrera, C.; Aime, S. Hydrogenative-PHIP Polarized Metabolites for Biological Studies. *Magn. Reson. Mater. Phy.* **2021**, *34* (1), 25–47. https://doi.org/10.1007/s10334-020-00904-x.

(25) Salnikov, O. G.; Chukanov, N. V.; Pravdivtsev, A. N.; Burueva, D. B.; Sviyazov, S. V.; Them, K.; Hövener, J.; Koptyug, I. V. Heteronuclear Parahydrogen-Induced Hyperpolarization via Side Arm Hydrogenation. *ChemPhysChem* **2025**, 2401119. https://doi.org/10.1002/cphc.202401119.

(26) Mamone, S.; Jagtap, A. P.; Korchak, S.; Ding, Y.; Sternkopf, S.; Glöggler, S. A Field-Independent Method for the Rapid Generation of Hyperpolarized [1-$^{13}$C]Pyruvate in Clean Water Solutions for Biomedical Applications. *Angew. Chem. Int. Ed.* **2022**, *61* (34), e202206298. https://doi.org/10.1002/anie.202206298.

(27) De Maissin, H.; Groß, P. R.; Mohiuddin, O.; Weigt, M.; Nagel, L.; Herzog, M.; Wang, Z.; Willing, R.; Reichardt, W.; Pichotka, M.; Heß, L.; Reinheckel, T.; Jessen, H. J.; Zeiser, R.; Bock, M.; Von Elverfeldt, D.; Zaitsev, M.; Korchak, S.; Glöggler, S.; Hövener, J.; Chekmenev, E. Y.; Schilling, F.; Knecht, S.; Schmidt, A. B. In Vivo Metabolic Imaging of [1-$^{13}$C]Pyruvate-d$_3$ Hyperpolarized By Reversible Exchange With

Parahydrogen**. *Angew. Chem. Int. Ed.* **2023**, *62* (36), e202306654. https://doi.org/10.1002/anie.202306654.

(28) Ding, Y.; Korchak, S.; Mamone, S.; Jagtap, A. P.; Stevanato, G.; Sternkopf, S.; Moll, D.; Schroeder, H.; Becker, S.; Fischer, A.; Gerhardt, E.; Outeiro, T. F.; Opazo, F.; Griesinger, C.; Glöggler, S. Rapidly Signal-enhanced Metabolites for Atomic Scale Monitoring of Living Cells with Magnetic Resonance. *Chem. Methods* **2022**, *2* (7), e202200023. https://doi.org/10.1002/cmtd.202200023.

(29) Berner, S.; Schmidt, A. B.; Zimmermann, M.; Pravdivtsev, A. N.; Glöggler, S.; Hennig, J.; von Elverfeldt, D.; Hövener, J. SAMBADENA Hyperpolarization of $^{13}$C-Succinate in an MRI: Singlet-Triplet Mixing Causes Polarization Loss. *ChemistryOpen* **2019**, *8* (6), 728–736. https://doi.org/10.1002/open.201900139.

(30) Schmidt, A. B.; Berner, S.; Schimpf, W.; Müller, C.; Lickert, T.; Schwaderlapp, N.; Knecht, S.; Skinner, J. G.; Dost, A.; Rovedo, P.; Hennig, J.; Von Elverfeldt, D.; Hövener, J.-B. Liquid-State Carbon-13 Hyperpolarization Generated in an MRI System for Fast Imaging. *Nat. Commun.* **2017**, *8* (1), 14535. https://doi.org/10.1038/ncomms14535.

(31) Pravdivtsev, A. N.; Brahms, A.; Ellermann, F.; Stamp, T.; Herges, R.; Hövener, J.-B. Parahydrogen-Induced Polarization and Spin Order Transfer in Ethyl Pyruvate at High Magnetic Fields. *Sci. Rep.* **2022**, *12* (1), 19361. https://doi.org/10.1038/s41598-022-22347-1.

(32) Bussandri, S.; Buljubasich, L.; Acosta, R. H. Diffusion Measurements with Continuous Hydrogenation in PHIP. *J. Magn. Reson.* **2020**, *320*, 106833. https://doi.org/10.1016/j.jmr.2020.106833.

(33) Hövener, J.-B.; Chekmenev, E. Y.; Harris, K. C.; Perman, W. H.; Tran, T. T.; Ross, B. D.; Bhattacharya, P. Quality Assurance of PASADENA Hyperpolarization for 13C Biomolecules. *Magn. Reson. Mater. Phy.* **2009**, *22* (2), 123–134. https://doi.org/10.1007/s10334-008-0154-y.

(34) Goldman, M.; Jóhannesson, H.; Axelsson, O.; Karlsson, M. Design and Implementation of 13C Hyper Polarization from Para-Hydrogen, for New MRI Contrast Agents. *C. R. Chim.* **2005**, *9* (3–4), 357–363. https://doi.org/10.1016/j.crci.2005.05.010.

(35) Li, L. Z.; Kadlecek, S.; Xu, H. N.; Daye, D.; Pullinger, B.; Profka, H.; Chodosh, L.; Rizi, R. Ratiometric Analysis in Hyperpolarized NMR (I): Test of the Two-site Exchange Model and the Quantification of Reaction Rate Constants. *NMR Biomed.* **2013**, *26* (10), 1308–1320. https://doi.org/10.1002/nbm.2953.

(36) Coffey, A. M.; Kovtunov, K. V.; Barskiy, D. A.; Koptyug, I. V.; Shchepin, R. V.; Waddell, K. W.; He, P.; Groome, K. A.; Best, Q. A.; Shi, F.; Goodson, B. M.; Chekmenev, E. Y. High-Resolution Low-Field Molecular Magnetic Resonance Imaging of Hyperpolarized Liquids. *Anal. Chem.* **2014**, *86* (18), 9042–9049. https://doi.org/10.1021/ac501638p.

(37) Schmidt, A. B.; Zimmermann, M.; Berner, S.; De Maissin, H.; Müller, C. A.; Ivantaev, V.; Hennig, J.; Elverfeldt, D. V.; Hövener, J.-B. Quasi-Continuous Production of Highly Hyperpolarized Carbon-13 Contrast Agents Every 15 Seconds within an MRI System. *Commun. Chem.* **2022**, *5* (1), 21. https://doi.org/10.1038/s42004-022-00634-2.

(38) Nagel, L.; Gierse, M.; Gottwald, W.; Ahmadova, Z.; Grashei, M.; Wolff, P.; Josten, F.; Karaali, S.; Müller, C. A.; Lucas, S.; Scheuer, J.; Müller, C.; Blanchard, J.; Topping, G. J.; Wendlinger, A.; Setzer, N.; Sühnel, S.; Handwerker, J.; Vassiliou, C.; Van Heijster, F. H. A.; Knecht, S.; Keim, M.; Schilling, F.; Schwartz, I. Parahydrogen-Polarized [1-$^{13}$C]Pyruvate for Reliable and Fast Preclinical Metabolic Magnetic Resonance Imaging. *Adv. Sci.* **2023**, *10* (30), 2303441. https://doi.org/10.1002/advs.202303441.

(39) Ellermann, F.; Sirbu, A.; Brahms, A.; Assaf, C.; Herges, R.; Hövener, J.-B.; Pravdivtsev, A. N. Spying on Parahydrogen-Induced Polarization Transfer Using a Half-Tesla Benchtop MRI and Hyperpolarized Imaging Enabled by Automation. *Nat. Commun.* **2023**, *14* (1), 4774. https://doi.org/10.1038/s41467-023-40539-9.

(40) Hune, T. L. K.; Mamone, S.; Schmidt, A. B.; Mahú, I.; D'Apolito, N.; Wiedermann, D.; Brüning, J.; Glöggler, S. Hyperpolarized Multi-Organ Spectroscopy of Liver and Brain Using 1-13C-Pyruvate Enhanced via Parahydrogen. *Appl. Magn. Reson.* **2023**, *54* (11–12), 1283–1295. https://doi.org/10.1007/s00723-023-01578-z.

(41) Carr, H. Y.; Purcell, E. M. Effects of Diffusion on Free Precession in Nuclear Magnetic Resonance Experiments. *Phys. Rev.* **1954**, *94* (3), 630–638. https://doi.org/10.1103/PhysRev.94.630.

(42) Meiboom, S.; Gill, D. Modified Spin-Echo Method for Measuring Nuclear Relaxation Times. *Rev. Sci. Instrum.* **1958**, *29* (8), 688–691. https://doi.org/10.1063/1.1716296.

(43) Stevanato, G.; Ding, Y.; Mamone, S.; Jagtap, A. P.; Korchak, S.; Glöggler, S. Real-Time Pyruvate Chemical Conversion Monitoring Enabled by PHIP. *J. Am. Chem. Soc.* **2023**, *145* (10), 5864–5871. https://doi.org/10.1021/jacs.2c13198.

(44) *Online Materials Information Resource - MatWeb*. https://www.matweb.com/index.aspx (accessed 2025-05-02).

(45) Ratajczyk, T.; Gutmann, T.; Dillenberger, S.; Abdulhussaein, S.; Frydel, J.; Breitzke, H.; Bommerich, U.; Trantzschel, T.; Bernarding, J.; Magusin, P. C. M. M.; Buntkowsky, G. Time Domain Para Hydrogen Induced Polarization. *Solid State Nucl. Magn. Reson.* **2012**, *43–44*, 14–21. https://doi.org/10.1016/j.ssnmr.2012.02.002.

(46) Pravdivtsev, A. N.; Ivanov, K. L.; Yurkovskaya, A. V.; Vieth, H.-M.; Sagdeev, R. Z. New Pulse Sequence for Robust Filtering of Hyperpolarized Multiplet Spin Order. *Dokl. Phys. Chem.* **2015**, *465* (1), 267–269. https://doi.org/10.1134/S0012501615110044.

(47) Pravdivtsev, A. N.; Sönnichsen, F.; Hövener, J.-B. OnlyParahydrogen SpectrosopY (OPSY) Pulse Sequences – One Does Not Fit All. *J. Magn. Reson.* **2018**, *297*, 86–95. https://doi.org/10.1016/j.jmr.2018.10.006.

(48) Eills, J.; Cavallari, E.; Kircher, R.; Di Matteo, G.; Carrera, C.; Dagys, L.; Levitt, M. H.; Ivanov, K. L.; Aime, S.; Reineri, F.; Münnemann, K.; Budker, D.; Buntkowsky, G.; Knecht, S. Singlet-Contrast Magnetic Resonance Imaging: Unlocking Hyperpolarization with Metabolism**. *Angew. Chem. Int. Ed.* **2021**, *60* (12), 6791–6798. https://doi.org/10.1002/anie.202014933.

(49) Kozinenko, V. P.; Kiryutin, A. S.; Knecht, S.; Buntkowsky, G.; Vieth, H.-M.; Yurkovskaya, A. V.; Ivanov, K. L. Spin Dynamics in Experiments on Orthodeuterium Induced Polarization (ODIP). *J. Chem. Phys.* **2020**, *153* (11), 114202. https://doi.org/10.1063/5.0022042.

(50) Barskiy, D. A.; Kovtunov, K. V.; Gerasimov, E. Y.; Phipps, M. A.; Salnikov, O. G.; Coffey, A. M.; Kovtunova, L. M.; Prosvirin, I. P.; Bukhtiyarov, V. I.; Koptyug, I. V.; Chekmenev, E. Y. 2D Mapping of NMR Signal Enhancement and Relaxation for Heterogeneously Hyperpolarized Propane Gas. *J. Phys. Chem. C* **2017**, *121* (18), 10038–10046. https://doi.org/10.1021/acs.jpcc.7b02506.

(51) Dagys, L.; Korzeczek, M. C.; Parker, A. J.; Eills, J.; Blanchard, J. W.; Bengs, C.; Levitt, M. H.; Knecht, S.; Schwartz, I.; Plenio, M. B. Robust Parahydrogen-Induced Polarization at High Concentrations. *Sci. Adv.* **2024**, *10* (30), eado0373. https://doi.org/10.1126/sciadv.ado0373.

(52) Sengstschmid, H.; Freeman, R.; Barkemeyer, J.; Bargon, J. A New Excitation Sequence to Observe the PASADENA Effect. *J. Magn. Reson. Ser. A.* **1996**, *120* (2), 249–257. https://doi.org/10.1006/jmra.1996.0121.

(53) *Hydrogen and Deuterium*, 1. ed.; Young, C. L., Ed.; Solubility data series; Pergamon Press: Oxford, 1981.

(54) Mohiuddin, O.; De Maissin, H.; Pravdivtsev, A. N.; Brahms, A.; Herzog, M.; Schröder, L.; Chekmenev, E. Y.; Herges, R.; Hövener, J.-B.; Zaitsev, M.; Von Elverfeldt, D.; Schmidt, A. B. Rapid in Situ Carbon-13 Hyperpolarization and Imaging of Acetate and Pyruvate Esters without External Polarizer. *Commun. Chem.* **2024**, *7* (1), 240. https://doi.org/10.1038/s42004-024-01316-x.

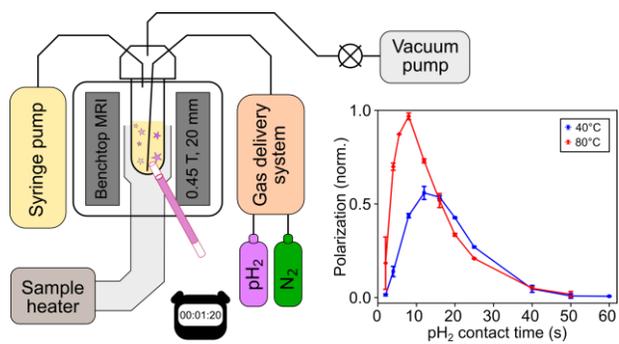

Supplementary Information

# Upscaling the hyperpolarization sample volume of an automated hydrogenative parahydrogen-induced polarizer

Yenal Gökpek[a], Jan-Bernd Hövener*[a], Andrey N. Pravdivtsev*[a]

[a] Section Biomedical Imaging, Molecular Imaging North Competence Center (MOIN CC), Department of Radiology and Neuroradiology, University Hospital Schleswig-Holstein, Kiel University, Am Botanischen Garten 14, 24114, Kiel, Germany

## Contents

# Simulations of nutation curves of the $B_1$ coil with different sample heights

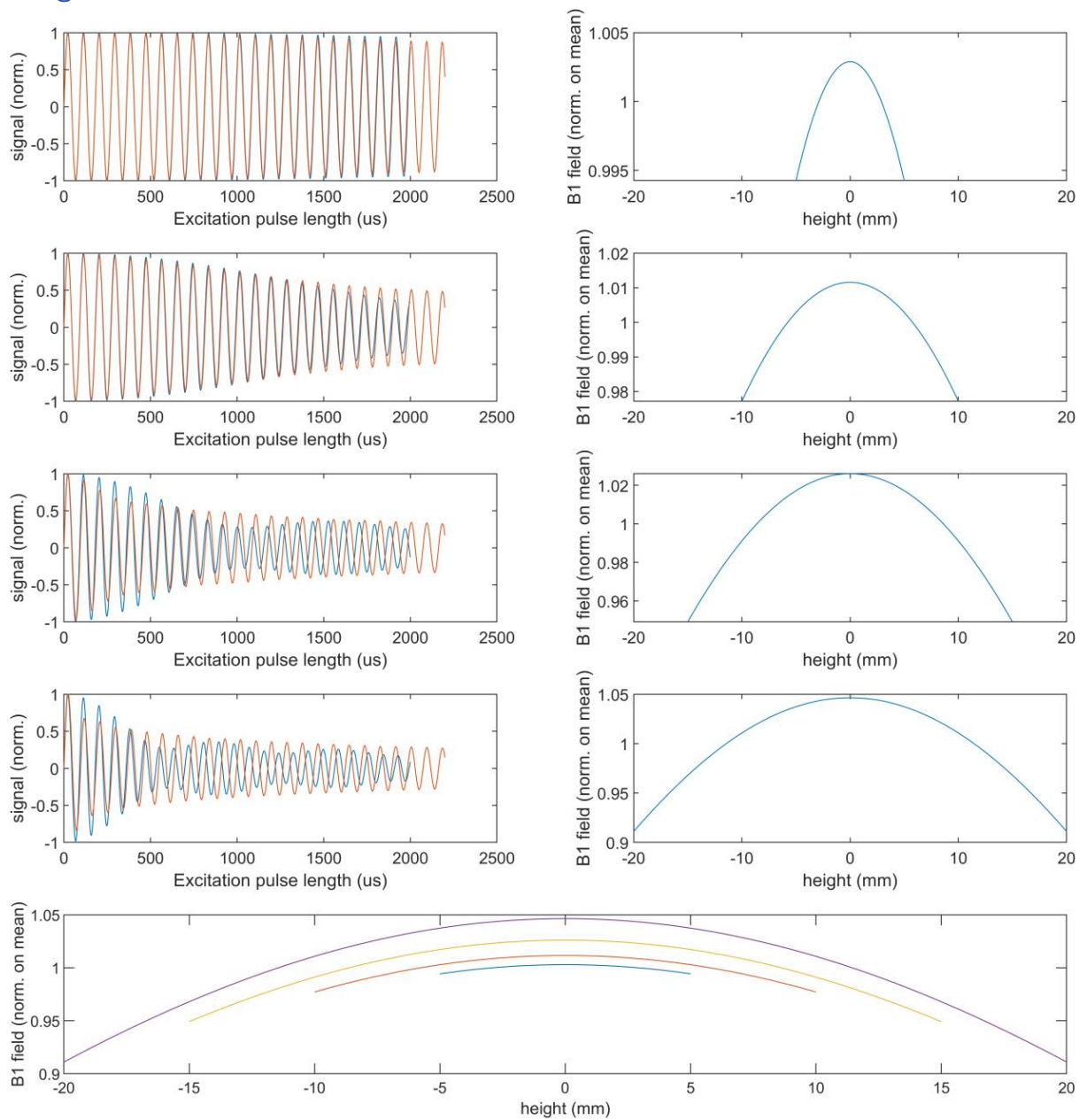

Figure S1. Simulated nutation decay parameters for 10, 20, 30, and 40 mm high acetone in a 16 mm reactor.

# Technical drawing of the 5 mm spinnerless cap with polymer tube

Figure S2. 5 mm spinnerless cap with polymer tube technical drawing

## Technical drawing of the 5 mm pressure cap

Figure S3. 5 mm pressure cap technical drawing

# Technical drawing of the 10 mm spinnerless cap with polymer tube

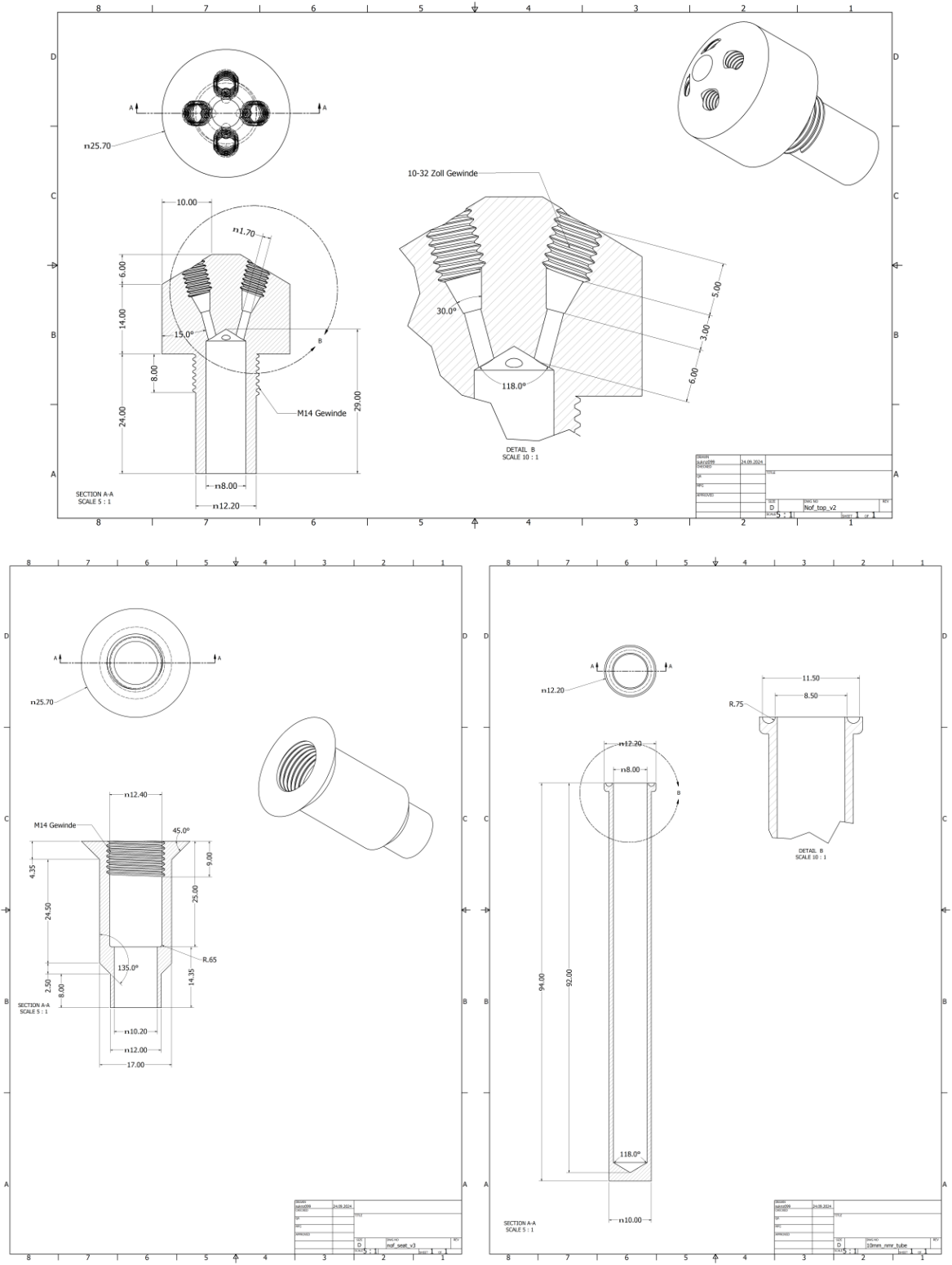

Figure S4. 10 mm spinnerless cap with polymer tube technical drawing

# Technical drawing of the 10 mm pressure cap

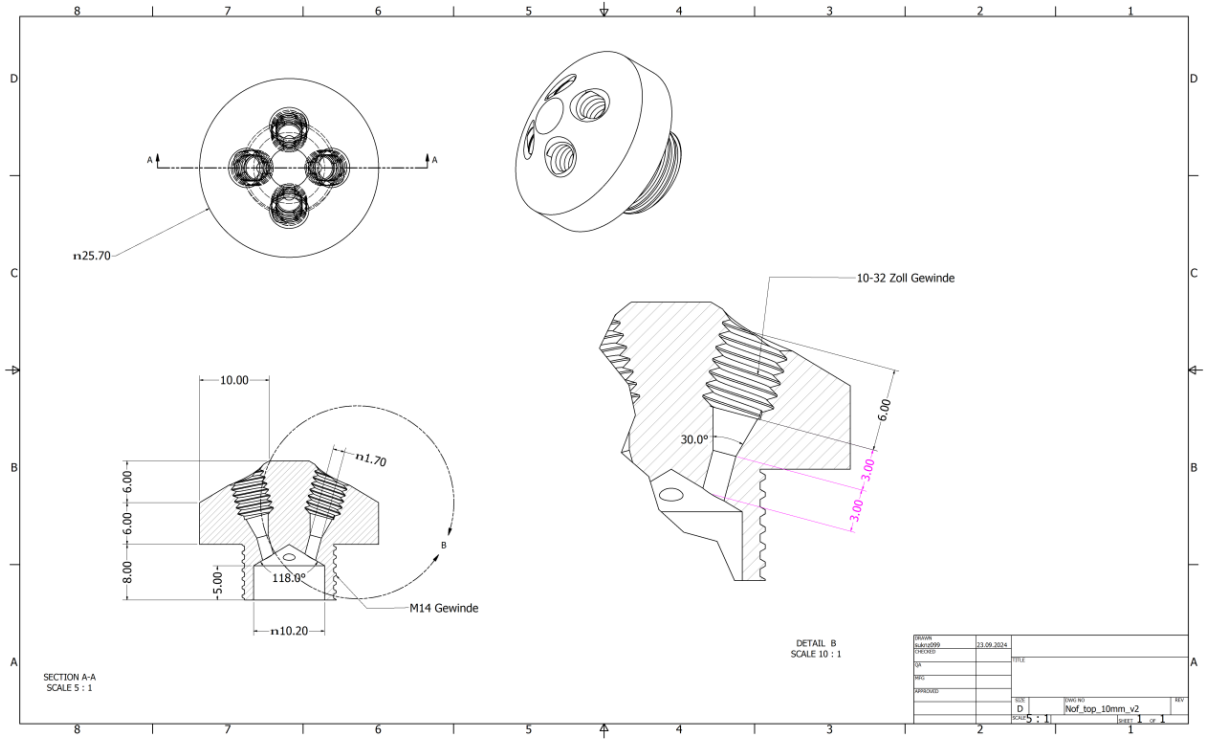

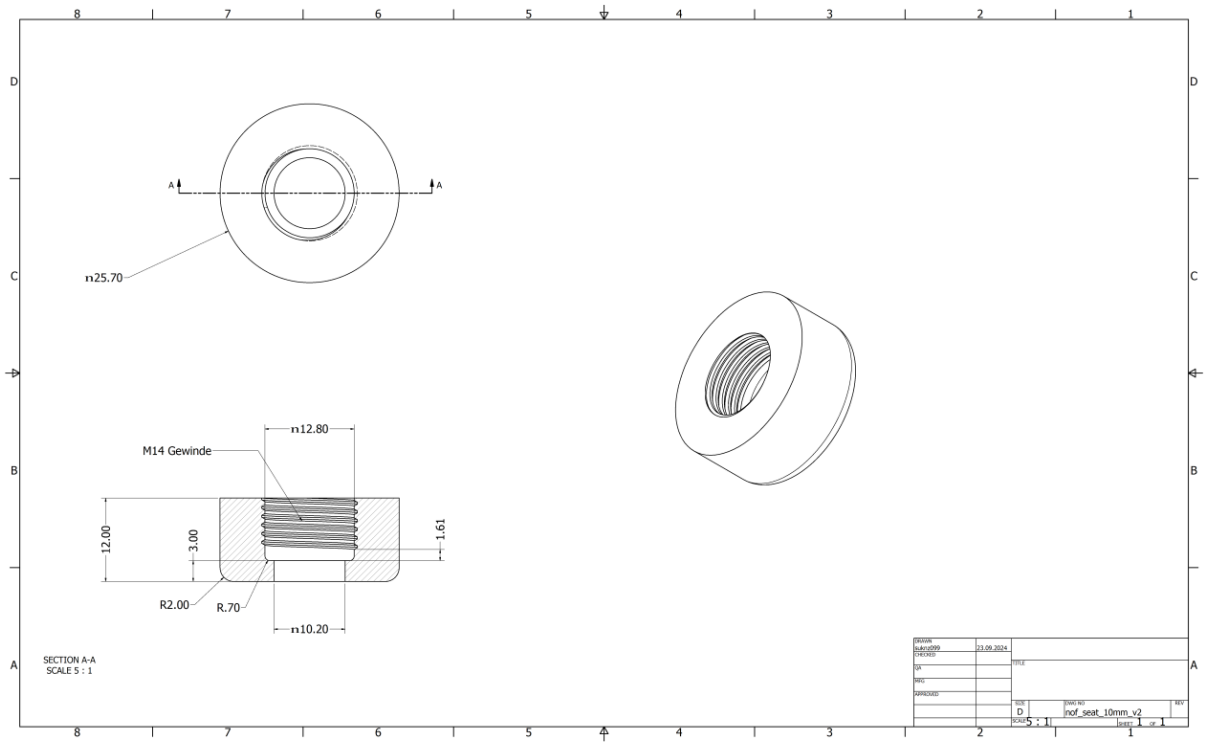

Figure S5. 10 mm pressure cap technical drawing

## Technical drawing of the 16 mm pressure cap

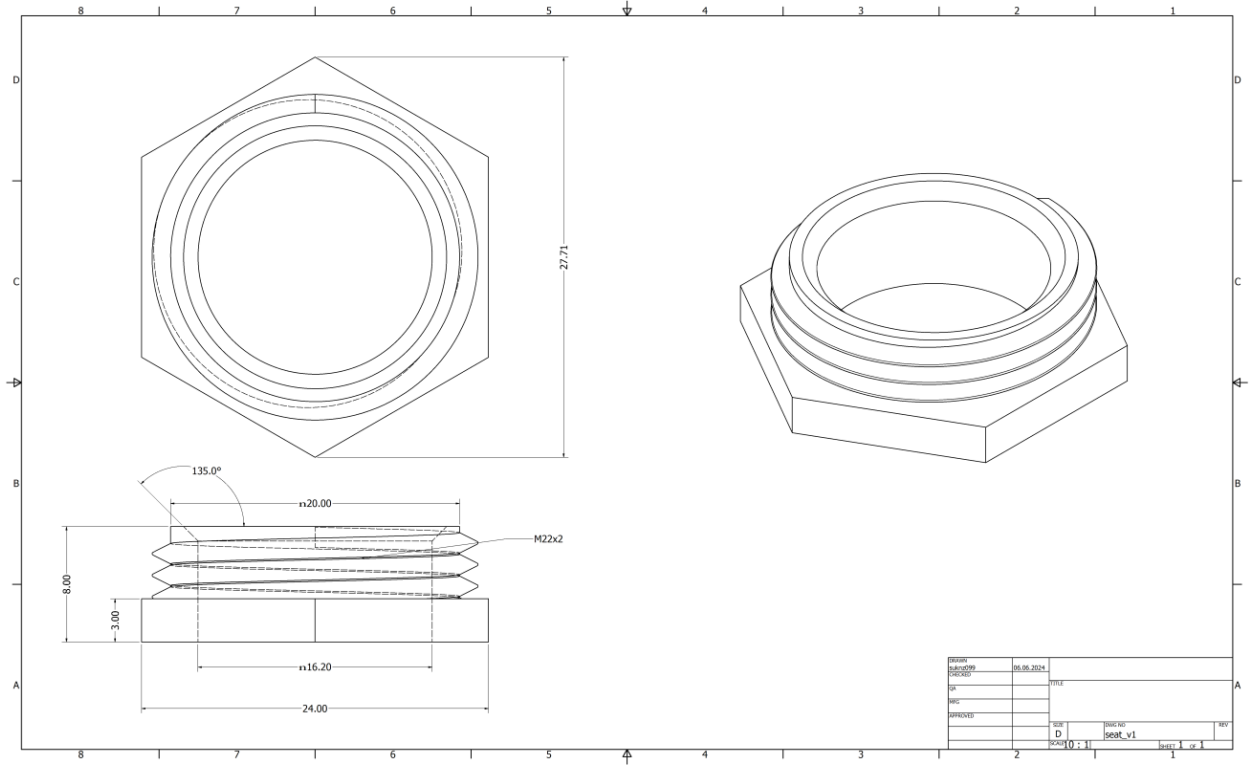

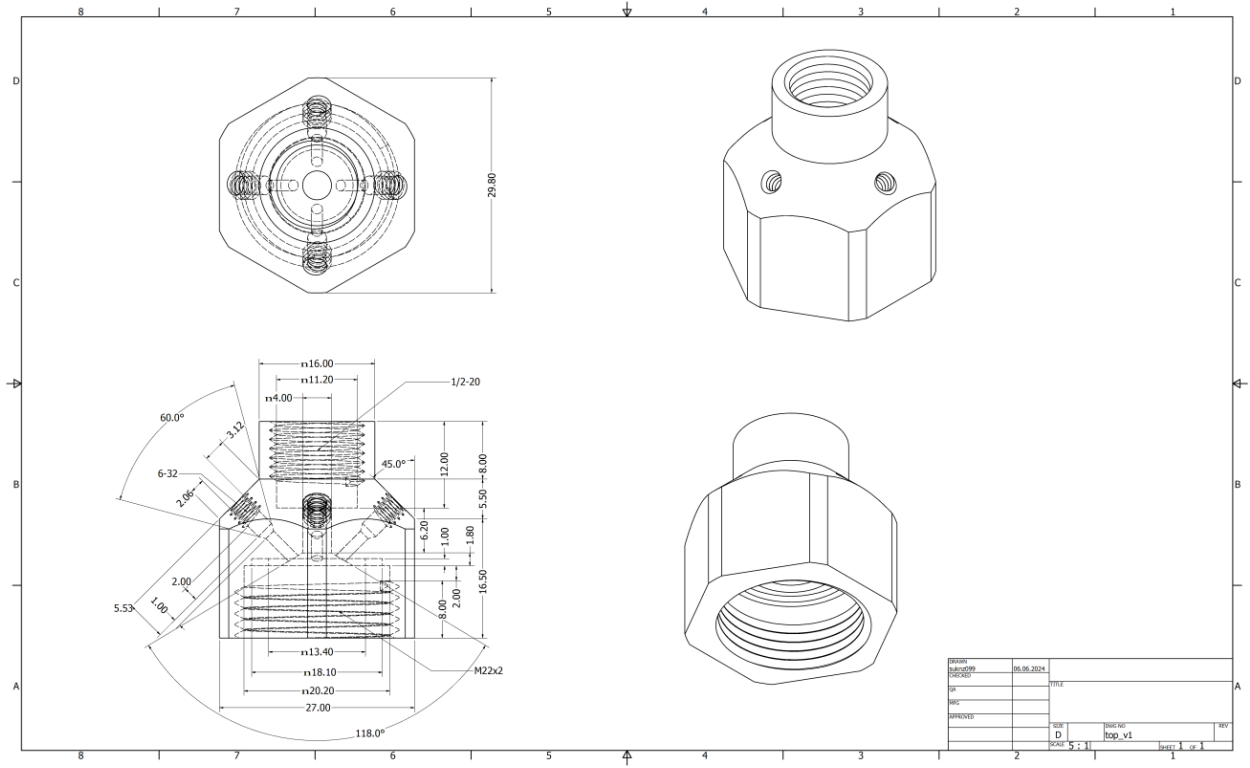

Figure S6. 16 mm pressure cap technical drawing

## Spectra of the product with high-resolution NMR

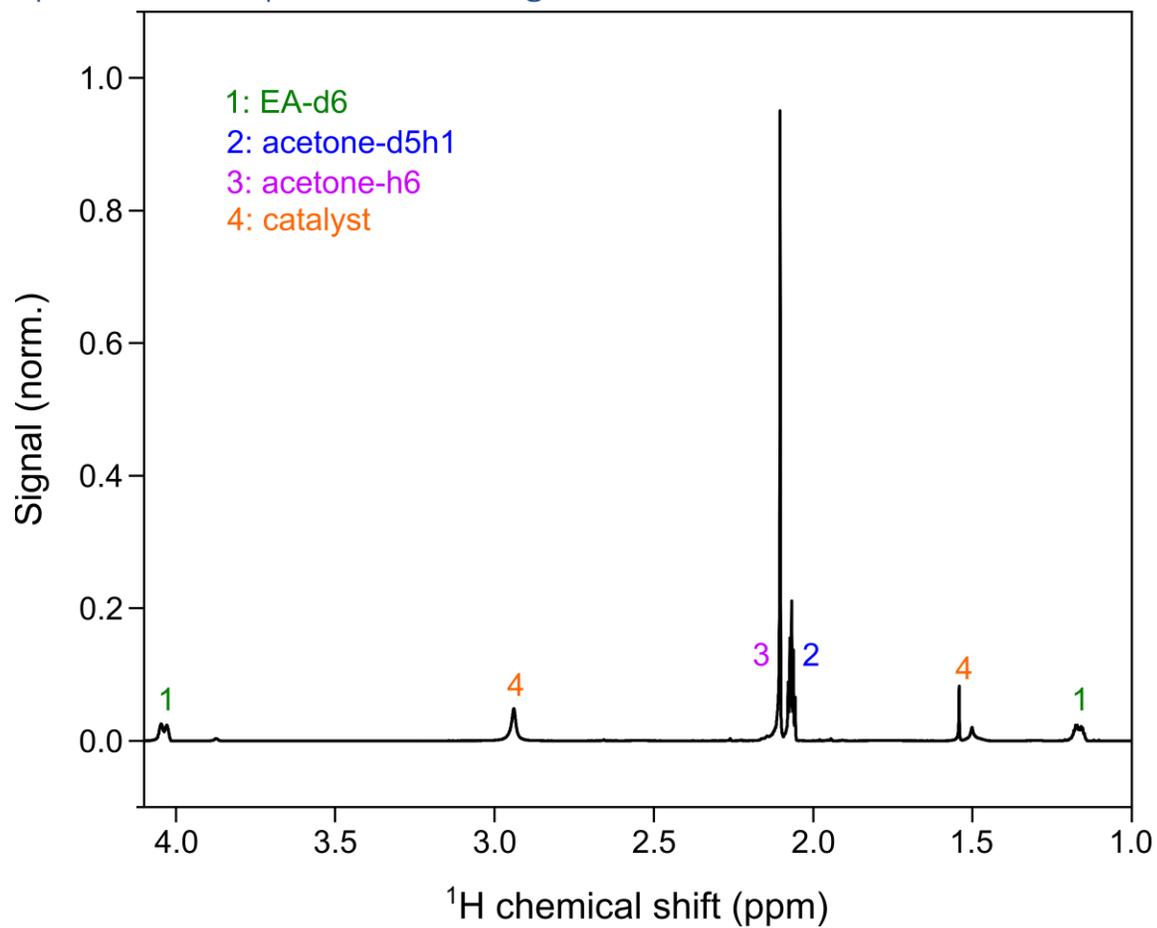

Figure S7. High resolution spectrum of the product